\documentclass[twocolumn]{aastex6}

\usepackage{graphicx}
\usepackage{dcolumn}
\usepackage{bm}
\usepackage{comment}
\usepackage{dcolumn}   
\usepackage{amssymb}   
\usepackage{amsmath}
\usepackage{tabularx}
\usepackage{hyperref}
\usepackage{xcolor}
\usepackage{tabulary}

\makeatletter
\def\mathcolor#1#{\@mathcolor{#1}}
\def\@mathcolor#1#2#3{%
  \protect\leavevmode
  \begingroup
    \color#1{#2}#3%
  \endgroup
}
\makeatother

\begin{document}

\preprint{APS/123-QED}

  \title{A triple origin for the heavy and low-spin binary black holes detected by LIGO/VIRGO}

\author{Carl L.\ Rodriguez\altaffilmark{1} and Fabio Antonini\altaffilmark{2}}

\altaffiltext{1}{Pappalardo Fellow; MIT-Kavli Institute for Astrophysics and Space Research, 77 Massachusetts Avenue, 37-664H, Cambridge, MA 02139, USA}
\altaffiltext{2}{Astrophysics Research Group, Faculty of Engineering and Physical Sciences,
University of Surrey, Guildford, Surrey, GU2 7XH, United Kingdom}


\date{\today}

\begin{abstract}
We explore the masses, merger rates, eccentricities, and spins
for field binary black holes driven to merger by a third companion through the
Lidov-Kozai mechanism.  Using a
population synthesis approach, we model the creation of
stellar-mass black hole triples across a range of different initial conditions 
and stellar metallicities.  We find that the production of triple-mediated 
mergers is enhanced at low metallicities by a factor of $\sim$ 100 due to the 
lower black hole natal kicks and reduced stellar mass loss.
These triples naturally yield heavy binary black holes with near-zero effective spins, consistent with most of the mergers observed to date.
This process produces a merger rate of between
2 and 25 $\rm{Gpc}^{-3} \rm{yr}^{-1}$ in the local universe, suggesting that the
Lidov-Kozai mechanism can potentially explain all of the low-spin, heavy black
hole mergers observed by Advanced LIGO/Virgo.  Finally, we show that triples
admit a unique eccentricity and spin distribution that will allow this model to
be tested in the near future.
\end{abstract}

\maketitle



\section{Introduction}
\label{sec:intro}
With the detection of 5 binary black hole (BH) mergers and one binary BH (BBH) candidate \citep{Abbott2016,Abbott2016a,Abbott2017,Abbott2017e,Abbott2017c},  Advanced LIGO and Advanced Virgo will soon begin providing entire catalogs of BBH mergers.  Even before the detection of GW150914, the first BBH merger, many different formation scenarios for BBHs had been proposed in the literature.   These include formation from isolated binaries, either through a common-envelope phase \citep[e.g.,][]{Belczynski2002,Voss2003,Podsiadlowski2003,Sadowski2007a,Belczynski2010,Dominik2012,Dominik2014,Dominik2013,Belczynski2016} or through chemically-homogeneous evolution via rapid rotation \citep[e.g.,][]{DeMink2016,Mandel2016a,Marchant2016}, dynamical formation in dense star clusters such as open clusters \citep[e.g.,][]{PortegiesZwart2000,Banerjee2010,Ziosi2014,Banerjee2017},
globular clusters \citep[e.g.,][]{PortegiesZwart2000,OLeary2006,OLeary2007,Moody2009,Downing2010,Downing2011,Tanikawa2013,Bae2014,Rodriguez2015a,Rodriguez2016a,Rodriguez2016b,Askar2016,Giesler2017,Rodriguez2018}, or galactic nuclei \citep[e.g.,][]{Miller2009,Oleary2009,Antonini2012a,Antonini2016,Bartos2016,Stone2016,VanLandingham2016,Leigh2017,Petrovich2017,Hoang2018}.  Despite their vastly different physical mechanisms, each of these formation channels have been invoked to solve the same problem: getting two black holes sufficiently close that the emission of gravitational waves (GWs) will lead them to merge.  

To solve this problem, \cite{Silsbee2017} and
\cite{Antonini2017} proposed an alternative solution which invokes
the secular interaction of a BBH with a third distant companion in the field of a galaxy.
  This third object can, at an appropriate separation and inclination, induce highly-eccentric oscillations in the BBH, which will in turn promote a rapid merger of the binary through GW emission.  This application of the \cite{Lidov1962} \cite{Kozai1962} (LK) mechanism 
 \citep[see][for a review]{Naoz2016}  provides a natural, purely dynamical mechanism to drive BBHs to merge in the field of a galaxy without having to invoke the complicated and poorly constrained physics of common-envelope evolution. 
  Despite the high multiplicity of triple systems around massive binaries \cite[$\sim 60\%$,][]{Sana2014}, the contribution of stellar triples to the BBH merger rate remains minimally explored.


Furthermore, it has been recently shown \citep{Liu2017,Antonini2017a,Liu2018} that the precession of the intrinsic spins of the BHs about the orbital angular momentum of the binary can produce significant misalignment between the orbital and spin angular momenta of merging BBHs from the LK channel.  In \cite{Antonini2017a}, we showed that the spin evolution of a BBH during LK oscillations naturally leads to effective spins near zero, consistent with many of the LIGO/Virgo detections to date.  Given that the spins of merging BBHs have been proposed as a promising way to discriminate between formation channels \citep{Gerosa2013,Rodriguez2016c,Vitale2017,Farr2017,Farr2018}, understanding the spin dynamics of any given formation channel is necessary to understand its contribution to the GW landscape. 

In this paper, we explore BBH mergers from binaries driven to merger by the LK effect in galactic fields.  We evolve a set of stellar triples to provide a realistic population of stellar-mass BH triples, which are in turn integrated using the secular LK equations including the relativistic spin-orbit (SO) and spin-spin (SS) couplings from post-Newtonian (pN) theory.   We find that at low metallicities the production of LK-driven mergers from stellar triples is significantly enhanced, largely due to the lower BH natal kicks and reduced mass lost to stellar winds.  Combined with an integration over the cosmic star formation rate, this enhancement suggests a BH merger rate between
1 and 25 $\rm{Gpc}^{-3} \rm{yr}^{-1}$ in the local universe, competitive with other formation channels.  These low-metallicity LK-driven mergers, with their large masses and near-zero effective spins, can easily explain all of the heavy, low-spin BBH mergers observed by LIGO/Virgo.

In Section \ref{sec:eom}, we re-derive the relativistic corrections to the 
binary motion arising from the precession of the pericenter and the lowest-order 
SO and SS terms, using a  Hamiltonian formalism developed in 
\cite{Tremaine2009,Petrovich2015,Liu2015}, and explore the resultant 
implications for the spin evolution.  In Section \ref{sec:pop}, we describe the 
setup of our triple population synthesis technique, while in Section 
\ref{sec:popsynth} we describe the features of our population of evolved BH 
triples.  In Section \ref{sec:spin}, we show how the distribution of BH spins 
from merging triples (and in particular the distributions of the effective 
spins) naturally forms a population of mergers with near-zero effective spins, 
while in Section \ref{sec:observable} we showcase various observable parameters 
from our BBH merger population (including the masses, eccentricities, and 
spins), and compute the merger rate of BH triples.  Throughout this paper, we 
assume a $\Lambda$CDM cosmology with $h= 0.679$ and $\Omega_M = 0.3065$ 
\citep{PlanckCollaboration2015}, {and that all BHs are born maximally spinning 
(although we relax this assumption in Section \ref{subsec:massandspins}}).

\section{Secular Equations of Motion}
\label{sec:eom}

\begin{figure*}[bth]
\centering
\includegraphics[scale=0.8]{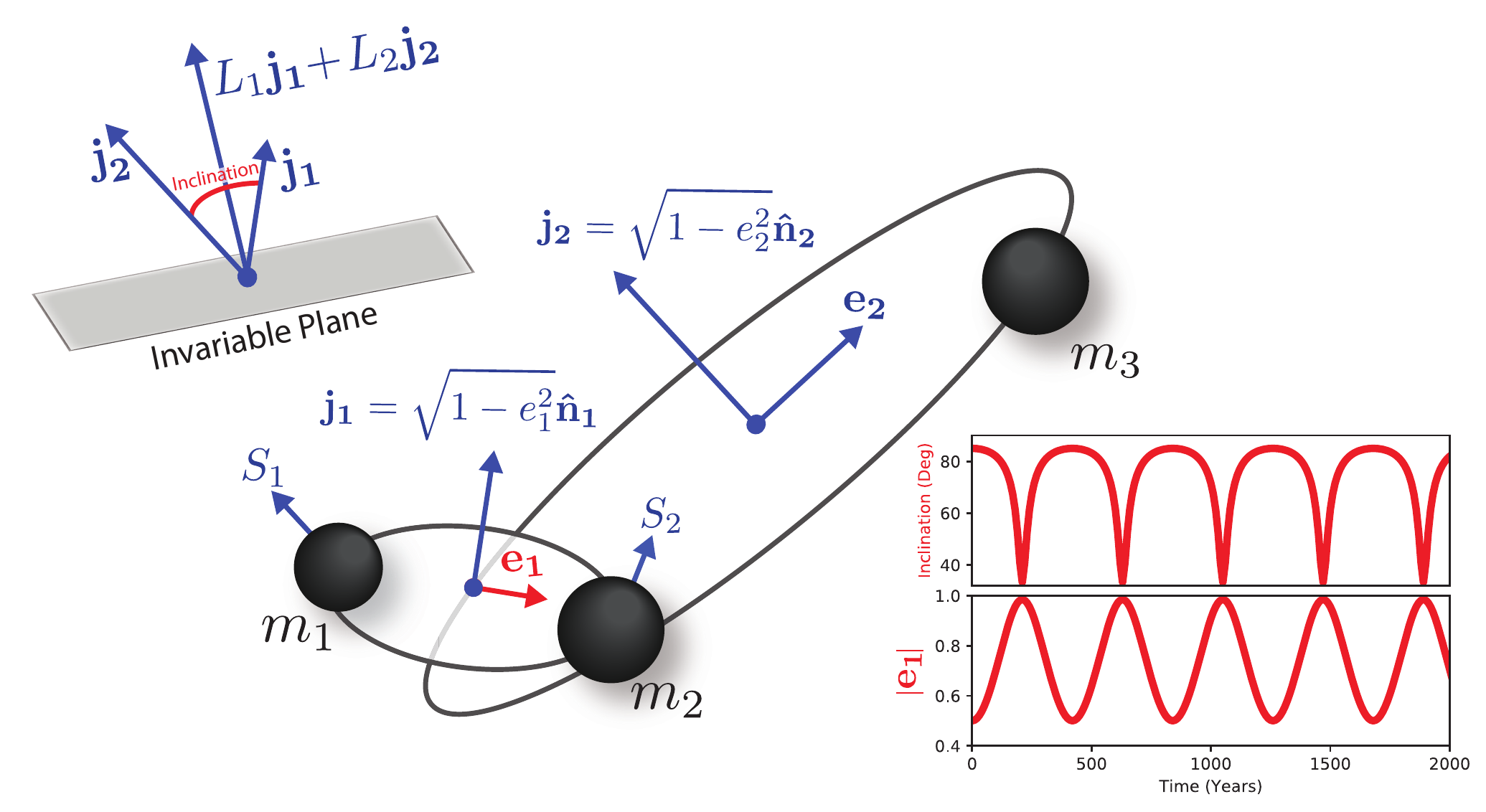}
\caption{An illustration of the general problem considered in this paper, and the geometric formalism we use.  For any triple system, the total angular momentum of the system is conserved (in the absence of GW emission), and defines a fixed ``invariable plane'' of the system.  $L_1$ and $L_2$ define the circular angular momenta of each binary (e.g. $L_1 = \mu \sqrt{G(m_1+m_2)a}$), while the dimensionless angular momentum ($\mathbf{j_{1}}$ and $\mathbf{j_{2}}$) and eccentricity ($\mathbf{e_{1}}$ and $\mathbf{e_{2}}$) vectors are used to define the orientation and orbital elements of the triple.   The eccentricity of the inner binary ($e_1$) and the mutual inclination of the two binaries change during an LK oscillation, as the two binaries exchange angular momentum while precessing about the total angular momentum of the system.  The plot shows several LK oscillations for a triple with $m_1=m_2=m_3=1M_{\odot}$, $a_1 = 1$ AU, $a_2 = 10$AU, $e_2 = 0.2$, and an initial state of $e_1 = 0.5$ and $i=85^{\circ}$.  $S_1$ and $S_2$ define the spin vectors of the inner BHs, which we assume do not directly couple to the outer binary.}
\label{fig:triple}
\end{figure*}

We are interested in the long-term evolution of triple systems for which the relativistic contributions to the inner binary become significant.  To set up our dynamical problem, we use the LK equations of motion to octupole order \citep{Petrovich2015,Liu2015} using the geometric formalism developed in \cite{Tremaine2009,Correia2011}.  See \cite{Tremaine2014} for a detailed explanation.  In this formalism, the orientations of the binaries and their orbital elements are described using the dimensionless angular momentum and Laplace-Runge-Lenz vectors, $\mathbf{j}$ and $\mathbf{e}$, defined such that

\begin{align}
\mathbf{j} &\equiv \sqrt{1 - e^2}\hat{\mathbf{n}}\nonumber\\
\mathbf{e} &\equiv e \hat{\mathbf{u}}\nonumber
\end{align}

\noindent where $e$ is the eccentricity of the binary, $\mathbf{e}$ points in the direction of the binary pericenter, and $\mathbf{j}$ points along the orbital angular momentum of the binary.  See Figure \ref{fig:triple}.  We also define the scalar angular momentum for a circular binary

\begin{equation*}
L \equiv \mu \sqrt{G M a}
\end{equation*}

\noindent such that $L \times \mathbf{j}$ is the standard angular momentum vector, with $M \equiv m_1 + m_2$ and $\mu \equiv m_1m_2 / M$ being the reduced mass of the binary.  

The power of this formalism lies in the fact that the Poisson brackets of the angular momentum and eccentricity vectors can be expressed as

\begin{align}
\{j_i,j_j\} = \frac{1}{L}\epsilon_{ijk}j_k \nonumber \\
\{e_i,e_j\} = \frac{1}{L}\epsilon_{ijk}j_k \nonumber \\
\{j_i,e_j\} = \frac{1}{L}\epsilon_{ijk}e_k \label{eqn:brackets}
\end{align}

\noindent where $\epsilon_{ijk}$ is the Levi-Civita symbol.  It is then straightforward to show that, for any Hamiltonian $H$, the equations of motion can be expressed as \citep{Tremaine2009}

\begin{align*}
\frac{df}{dt} &= \{f,H\} \\
&= \{f,\mathbf{j}\}\mathbf{\nabla}_{\mathbf{j}}H + \{f,\mathbf{e}\}\mathbf{\nabla}_{\mathbf{e}}H\\
\end{align*}

\noindent which, when combined with the Poisson brackets from Eqn.~\eqref{eqn:brackets} yields

\begin{align}
\frac{d\mathbf{j}}{dt} &= \frac{1}{L} \left( \mathbf{e} \times \mathbf{\nabla}_{\mathbf{e}}H + \mathbf{j} \times \mathbf{\nabla}_{\mathbf{j}}H\right) \label{eqn:djdt}\\
\frac{d\mathbf{e}}{dt} &= -\frac{1}{L} \left( \mathbf{j} \times \mathbf{\nabla}_{\mathbf{e}}H + \mathbf{e} \times \mathbf{\nabla}_{\mathbf{j}}H\right) \label{eqn:dedt} .
\end{align}

\noindent To determine the orbital evolution of a system, all that remains is to specify the orbit-averaged Hamiltonian to be inserted into Equations \eqref{eqn:djdt} and \eqref{eqn:dedt}.

For a non-spinning triple system, the LK Hamiltonian $H_{\rm LK}$ can be written as 

\begin{equation}
H_{\rm{LK}} = H_{1} + H_{2} + H_{12}
\end{equation}

\noindent where $H_{1}$ and $H_{2}$ are the Keplerian Hamiltonians for the inner and outer binaries, and $H_{12}$ is an interaction term between the two, often expressed as a series expansion in the instantaneous separation between the two binaries \citep{Harrington1968,Kozai1962,Lidov1962,Ford2000,Naoz2013}. We use the geometric form of these equations developed in \cite{Petrovich2015} and \cite{Liu2015}, accurate up to the octupole order of the interaction term.  See  Equations (17)-(20) in \cite{Liu2015}.

To account for the pN corrections to the Newtonian three-body problem, we can self-consistently add additional terms to the Hamiltonian, accounting for pericenter precession at first pN order (1pN), the evolution of the BH spins due to geodedic precession about $\mathbf{j_1}$ (1pN), the back-reaction on the orbit from Lense-Thirring precession \citep[1.5pN,][]{Damour1988}, and the gravitomagnetic coupling between the two spins and the quadrupole-monopole interaction \citep[2pN,][]{Barker1975,Damour2001}.   The complete Hamiltonian then becomes: 

\begin{equation}
H = H_{\rm{LK}} + H_{1\rm{pN}} + H_{\rm{SO}} + H_{\rm{SS}}
\label{eqn:hamil}
\end{equation}

\noindent where the pN coupling between the orbit of the inner masses and their spins are given by \citep[e.g.,][]{Damour1988,Buonanno2011}

\begin{align}
H_{1\rm{pN}} &= \frac{\mu}{c^2} \left[ \frac{(3\mu/M-1)\mathbf{p_1}^4}{8} - \frac{G\mu(\mathbf{p_1}\cdot\mathbf{\hat{r}_1})^2}{2r_1} \right. \nonumber \\
&~~~~\ \ \ \ - \left. \frac{(3+\mu/M)GM\mathbf{p_1}^2}{2r_1} + \frac{(GM)^2}{2r_1^2} \right] \\
H_{\rm{SO}} &= \frac{2 G L_1}{c^2|\mathbf{r_1}|^3} \mathbf{S_{\rm{eff}}}\cdot\mathbf{j_1}\\
H_{\rm{SS}} &= \frac{G \mu}{2 c^2 M |\mathbf{r_1}|^3} \left [3(\mathbf{S_0} \cdot \mathbf{\hat{n}_1})^2 - \mathbf{S_0}^2\right] 
\end{align}

\noindent where $\mathbf{r_1}$ and $\mathbf{p_1}$ are the reduced positions and momenta of the inner binary, and where we introduce two new combinations of the BH spin vectors:

\begin{align*}
\mathbf{S_{\rm{eff}}} &= \left(1+\frac{3m_2}{4m_1}\right)\mathbf{S_1} + \left(1+\frac{3m_1}{4m_2}\right)\mathbf{S_2}\\
\mathbf{S_{0}} &= \left(1+\frac{m_2}{m_1}\right)\mathbf{S_1} + \left(1+\frac{m_1}{m_2}\right)\mathbf{S_2}
\end{align*}

\noindent where $\boldsymbol{S_1}$ and $\boldsymbol{S_2}$ correspond to the spin vectors of $m_1$ and $m_2$.
These can also be written in terms
of the dimensionless spin parameter $\boldsymbol{\chi}$ as
\begin{align}\label{Equation:DefS}
\boldsymbol{S}_i = \boldsymbol{\chi}_i\frac{Gm_i^2}{c},\ \ \ \ 
|\boldsymbol{\chi}_i|\le 1.
\end{align}

Averaging each of these terms over an orbital period \cite[see e.g.][]{Tremaine2014}, we can express the orbit-averaged contributions from each pN effect as:

\begin{align}
\left<H_{1\rm{pN}}\right> &= \mathcolor{black}{\frac{G^2 M^2 \mu}{8 c^2 a_1^2} \left[ 15-\frac{\mu}{M} - \frac{24}{\sqrt{1-e_1^2}} \right]} \label{eqn:1pn}
\\
\left<H_{\rm{SO}}\right> &= \frac{2 G L_1}{c^2 a_1^3 (1-e_1^2)^{3/2}} \mathbf{S_{\rm{eff}}}\cdot\mathbf{j_1} \label{eqn:hso}
\\
\left<H_{\rm{SS}}\right> &= \frac{G \mu}{2 c^2 M a_1^3 (1-e_1^2)^{3/2}} \left [\frac{1}{2}\mathbf{S_0}^2\ - \frac{3}{2}(\mathbf{S_0} \cdot \mathbf{\hat{n}_1})^2\right] \label{eqn:hss} .
\end{align}

To derive the equations of motion from Eqns.~(\ref{eqn:1pn}-\ref{eqn:hss}), we simply calculate the derivatives using Equations\ \eqref{eqn:djdt} and \eqref{eqn:dedt}.  For the contribution from $\left<H_{1\rm{pN}}\right>$, we find:

\begin{align}
\frac{d\mathbf{e_1}}{dt}\Big|_{1\rm{pN}} &= \{ \mathbf{e_1} , \left<H_{1\rm{pN}}\right> \}\nonumber\\
&=-\frac{1}{L_1} \left[ \{\mathbf{e_1},\mathbf{e_1}\} {\nabla}_{\mathbf{e_1}}\left<H_{1\rm{pN}}\right> + \{\mathbf{e_1},\mathbf{j_1}\} {\nabla}_{\mathbf{j_1}}\left<H_{1\rm{pN}}\right> \right]\nonumber\\
 &= \frac{ 3 (GM)^{3/2}  }{c^2 a_1^{5/2} (1-e_1^2)^{3/2} } \mathbf{j_1}\times\mathbf{e_1}
\label{eqn:1pne1}
\end{align}

\noindent with $\mathbf{j_1}$ conserved at 1pN order prior to the inclusion of spin effects or GW emission.  This is identical to the pericenter precession term found in the literature \citep{Eggleton2001,Fabrycky2007,Liu2015}, but falls naturally out of the orbit-averaged vector formalism.  We then consider the SO and SS terms.  In addition to the Poisson brackets between $\mathbf{j}$ and $\mathbf{e}$, we introduce Poisson brackets for the spin vectors:

\begin{align}
\{S_1^i,S_1^j\} &= \epsilon_{ijk}S_1^i S_1^j\nonumber\\
\{S_2^i,S_2^j\} &= \epsilon_{ijk}S_2^i S_2^j\nonumber\\
\{S_1^i,S_2^j\} &= 0
\end{align}

\noindent with all other Poisson brackets ($\{S_1^i,j_1^j\}$, $\{S_1^i,e_1^j\}$, $\{S_1^i,j_2^j\}$, $\{S_1^i,e_2^j\}$, and their $1\leftrightarrow2$ equivalents) being zero.  We can then explicitly write down the orbit-averaged equations of motion from $\left<H_{\rm{SO}}\right>$ and $\left<H_{\rm{SS}}\right>$:

\begin{align}
\frac{d\mathbf{j_1}}{dt}\Big|_{\rm{SO}} &= \frac{2 G}{c^2 a_1^3 (1-e_1^2)^{3/2}} \mathbf{S_{\rm{eff}}}\times\mathbf{j_1}\label{eqn:djdtso}\\
\frac{d\mathbf{e_1}}{dt}\Big|_{\rm{SO}} &= \frac{2 G}{c^2 a_1^3 (1-e_1^2)^{3/2}} \left[\mathbf{S_{\rm{eff}}}-3(\mathbf{S_{\rm{eff}}}\cdot\mathbf{\hat{n}_1})\mathbf{\hat{n}_1}\right]\times \mathbf{e_1}\label{eqn:dedtso}\\
\frac{d\mathbf{S_1}}{dt}\Big|_{\rm{SO}} &= \frac{2 G^{3/2}\mu M^{1/2}}{c^2 a_1^{5/2} (1-e_1^2)^{3/2}} \left(1+\frac{3m_2}{4m_1}\right)\mathbf{j_1}\times \mathbf{S_1} \label{eqn:dsdtso}
\end{align}

\noindent and similarly for $\mathbf{S_2}$.  The SS terms can be derived in a similar fashion:

\begin{align}
\left.\frac{d\mathbf{j_1}}{dt}\right\rvert_{\rm{SS}} &= -\frac{3 G^{1/2}}{2c^2 M^{3/2} a_1^{7/2} (1-e_1^2)^{2}} (\mathbf{S_{\rm{0}}}\cdot\mathbf{\hat{n}_1})\mathbf{S_0}\times \mathbf{j_1}\label{eqn:djdtss}\\
\left.\frac{d\mathbf{e_1}}{dt}\right\rvert_{\rm{SS}} &= \frac{3 G^{1/2}}{4c^2 M^{3/2} a_1^{7/2} (1-e_1^2)^{2}} \big[5(\mathbf{S_{\rm{0}}}\cdot\mathbf{\hat{n}_1})^2\mathbf{\hat{n}_1}- \nonumber\\ &~~~~~~2(\mathbf{S_{\rm{0}}}\cdot\mathbf{\hat{n}_1})\mathbf{S_{\rm{0}}} - \mathbf{S_{\rm{0}}}^2 \mathbf{\hat{n}_1}\big]\times \mathbf{e_1}\label{eqn:dedtss}\\
\left.\frac{d\mathbf{S_1}}{dt}\right\rvert_{\rm{SS}} &= \frac{G \mu}{2 c^2 M a_1^3 (1-e_1^2)^{3/2}}\left(1+\frac{m_2}{m_1}\right)\times \nonumber\\ &~~~~~~\big[\mathbf{S_0} - 3(\mathbf{S_0}\cdot \mathbf{\hat{n}_1})\mathbf{\hat{n}_1} \big]\times \mathbf{S_1} \label{eqn:dsdtss}
\end{align}

\noindent and similarly for $\mathbf{S_2}$. Equations \eqref{eqn:1pne1} and (\ref{eqn:djdtso}-\ref{eqn:dsdtss}), combined with the octupole-order LK equations from \cite{Liu2015}, give us the complete equations of motion for a triple with spinning inner components described by the Hamiltonian in Equation \eqref{eqn:hamil}, and are fully consistent with previously derived results in the literature for isolated binaries \citep{Barker1975}.

When defining the BH spins, we find it convenient to define the angles between the spins and the orbital angular momentum as 

\begin{equation*}
\theta_{1} \equiv \cos^{-1}\left(\frac{\mathbf{S_{1}} \cdot \mathbf{j_1}}{\lvert \mathbf{S_{1}} \cdot \mathbf{j_1} \rvert}\right)
\end{equation*}

\noindent and similarly for $\theta_2$.  We also define the angle between the components of the spins lying in the orbital plane 

\begin{equation*}
\Delta \phi \equiv \cos^{-1} \left( \frac{\mathbf{j_1} \times \mathbf{S_1}}{\lvert \mathbf{j_1} \times \mathbf{S_1}\rvert} \cdot \frac{\mathbf{j_1} \times \mathbf{S_2}}{\lvert \mathbf{j_1} \times \mathbf{S_2}\rvert} \right).
\end{equation*}

\noindent However, the spin parameter best constrained by the current generation of GW detectors is the effective binary spin  \citep{Ajith2011,Vitale2014,Purrer2016}, defined as the mass-weighted projection of the two spins onto the orbital angular momentum:

\begin{equation}
\chi_{\rm{eff}} \equiv \left(\frac{m_1 \mathbf{\chi_1} + m_2 \mathbf{\chi_2}}{m_1 + m_2}\right) \cdot \mathbf{\hat{n}_1},
\label{eqn:chieff}
\end{equation}

\noindent In the following sections, we will primarily focus on $\chi_{\rm{eff}}$ as the main spin observable of interest to Advanced LIGO/Virgo.

Since each of the above dynamical equation resembles an expression for simple precession (e.g.\ $\mathbf{\dot{u}} = \mathbf{\Omega} \times \mathbf{u}$), it is straight-forward to write down the timescale associated with each effect as $t \approx \pi/|\mathbf{\Omega}|$.  At 1pN order, the timescale for the precession of $\mathbf{e_1}$ about $\mathbf{j_1}$ (Eqn.~\eqref{eqn:1pne1}, often referred to as pericenter, Schwarzschild, or apsidal precession) is:

\begin{equation}
t_{1\rm{pN}} = \frac{\pi c^2 a_1^{5/2}(1-e_1^2)}{3(G M)^{3/2}} .
\label{eqn:t1pn}
\end{equation}

\noindent Also at 1pN order is the timescale for the precession of the spins $\mathbf{S_1}$ and $\mathbf{S_2}$ about $\mathbf{j_1}$ is 

\begin{equation}
t_{\mathbf{S_1}} = \frac{\pi c^2 a_1^{5/2}(1-e_1^2)}{2G^{3/2}\mu M^{1/2}} \left(1+\frac{3m_2}{4m_1}\right)^{-1}
\label{eqn:t1pnspin}
\end{equation}

\noindent and similarly for $\mathbf{S_2}$.  Note that, while SO effects are frequently referred to as 1.5pN order, the \emph{simple precession of the spins} about $\mathbf{j_1}$ is formally a 1pN effect.  This effect, often referred to as de Sitter or geodedic precession, is nothing more than the change in the angles arising from the parallel transport of the spins about the orbit.  Similar timescales can be derived for the higher-order SO and SS effects. Although we do not write them down here, we note that the timescales from Equations (\ref{eqn:djdtso} - \ref{eqn:dsdtss}) agree with results in the literature \citep[e.g.,][]{Merritt2013} in the limit of $\mathbf{S_2}\rightarrow 0$.  Additionally, we define the quadrupole timescale for a single LK oscillation of the triple as:

\begin{equation}
t_{\rm{LK}} = \frac{m_1+m_2}{\nu m_3}\left(\frac{a_2 \sqrt{1-e^2_2}}{a_1}\right)^3 
\label{eqn:tkozai}
\end{equation}

\noindent where $\nu \equiv \sqrt{G (m_1 + m_2)/a_1^3}$ is the mean motion of the inner binary.

Finally, we add the dissipation in $a_1$ and $e_1$ from the emission of GWs.  As writing down Hamiltonians for non-conservative processes requires special mathematical care, we instead simply add the known contributions from \cite{Peters1964} 

\begin{align}
\left<\frac{da}{dt}\right> &= -\frac{64}{5} \frac{G^3 m_1 m_2 (m_1+m_2)}{c^5 a^3 (1-e^2)^{7/2}} \left( 1+\frac{73}{24}e^2 + \frac{37}{96}e^4 \right)\label{eqn:dadt}\\
\left<\frac{de}{dt}\right> &= -\frac{304}{15} e \frac{G^3 m_1 m_2 (m_1+m_2)}{c^5 a^4 (1-e^2)^{5/2}} \left( 1 + \frac{121}{304} e^2 \right) .\label{eqn:dedtgw}
\end{align}

\noindent During each integration timestep, we compute the change in $a_1$ from Eqn.~\eqref{eqn:dadt}, while the change in eccentricity is included in the geometric variables as:

\begin{align*}
\left.\frac{d\mathbf{e_1}}{dt}\right\rvert_{\rm{GW}} &= \left<\frac{de}{dt}\right> \mathbf{\hat{e}_1}\\
\left.\frac{d\mathbf{j_1}}{dt}\right\rvert_{\rm{GW}} &= -\left<\frac{de}{dt}\right>  \frac{e_1}{\sqrt{1-e_1^2}}\mathbf{\hat{n}_1} \\
\end{align*} 

\noindent allowing us to self-consistently track the change in angular momentum from GW emission during our triple integration.

\subsection{Secular Dynamics of Spinning Triples}

The chaotic evolution of spins during the evolution of triple systems has been well described in the context of planetary systems \citep[e.g.][]{Storch2014,Storch2015,Liu2015}.  In that case, the evolution of the spin vector can be classified by comparing the precession timescales of the orbital angular momentum of the binary (during an LK oscillation) to the precession of the spin vector about the total angular momentum of the inner binary (due to tidal forces).  If the precession rate of $\mathbf{S_{1}}$ about $\mathbf{j_1}$ is significantly longer than the precession of $\mathbf{j_1}$ about the total angular momentum of the system, then the spins are expected to effectively precess about the total angular momentum, keeping a constant angle with respect to the invariable plane (the ``non-adiabatic'' case).  For our problem, this would correspond to $t_{\mathbf{S_1}} > t_{\rm{LK}}$.   On the other hand, if the precession of $\mathbf{S_{1}}$ about $\mathbf{j_1}$ is significantly faster than the precession of $\mathbf{j_1}$ (i.e. $t_{\mathbf{S_1}} < t_{\rm{LK}}$), the spins are expected to adiabatically follow $\mathbf{j_1}$, virtually oblivious to the presence of the third companion.  The intermediate ``trans-adiabatic'' regime, in which $t_{\mathbf{S_1}} \sim t_{\rm{LK}}$, allows for chaotic evolution of the spins, and is thought to play an important role in the observed spins of many exo-planetary systems \citep{Storch2014}.  

\begin{figure*}[bt]
\centering
\includegraphics[scale=0.95]{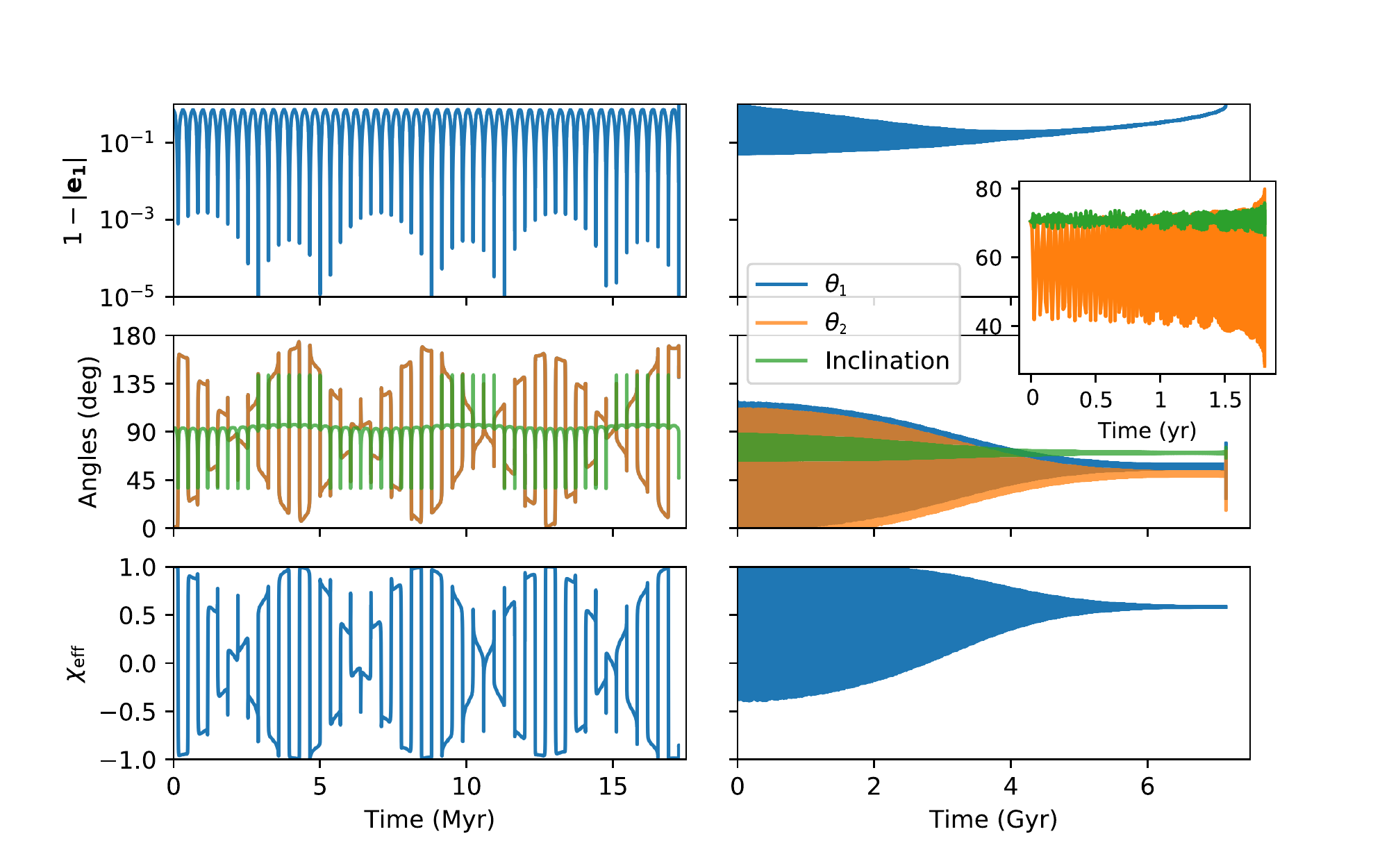}
\caption{Two triple systems integrated to merger from our BH triple population 
described in Section \ref{sec:popsynth}.  On the left, we show a triple where 
the eccentric LK mechanism pushes the inner binary to merge very rapidly (case 
i, from Section \ref{sec:2.5}).  {In this case, the spins remain 
essentially fixed in 3D space (with respect to the invariable plane) while the 
orbit of the inner binary oscillates wildly about them.}  On the right, we show an example where the spin precession and LK oscillations occur on a similar timescale.  Here, the spin vectors smoothly precess onto the invariable plane of the triple \citep[see][]{Antonini2017a}, until both they and the inner binary are frozen at a given orientation due to the increased pericenter precession from the 1pN terms (case ii).  Eventually this frozen binary merges due to GW emission.  The insert shows the final changes in inclination and $\theta_2$ arising from SO and SS effects, where the spins of the inner binary can couple to its orbital angular momentum (influencing the mutual inclination of the triple).  These effects are not present in $\chi_{\rm eff}$, since the effective spin is a constant of the motion up to 2pN order.}
\label{fig:examples}
\end{figure*}

Although we might expect to observe a similar behavior for the spinning BH systems studied here, we now show that this is not the case. We define the adiabaticity  parameter (with $\mathbf{S_1}$ as the spin of the more massive BH) as:

\begin{equation}
R \equiv \frac{t_{\rm{LK}}}{t_{\mathbf{S_1}}}
\label{eqn:adiab}
\end{equation}

\noindent such that any triple for which $R<1$ is considered non-adiabatic, while those where $R>1$ are considered adiabatic. While the LK timescale remains unchanged during the evolution of the triple (ignoring GW emission), the timescale for spin precession goes as ($1-e_1^2$),  meaning that the spin timescale could decrease by several orders of magnitude over the course of a single LK oscillation.  Thus, one might think that
a triple can easily transit  from the non-adiabatic to the adiabatic regime, where its brief moment in the trans-adiabatic regime can induce chaotic evolution in the spin vectors. 

Unfortunately, the potential chaotic spin evolution is largely suppressed by a conspiracy of relativity \citep{Antonini2017a}.   Both the precession of $\mathbf{S}_{1}$, $\mathbf{S}_2$ about $\mathbf{j_1}$ (Equation \ref{eqn:t1pnspin}) and the precession of the pericenter of the inner binary (Equation \ref{eqn:t1pn}) are formally 1pN order contributions, and both scale as $a_1^{5/2}(1-e_1^2)$.  However, it is well known that short range forces between components of the inner binary--particularly pericenter precession--can quench LK oscillations in a hierarchical triple, since at a certain separation the inner binary will precess so rapidly it decouples from the outer binary.  Conceptually, this point can be thought of as the separation where the time for the inner binary to precess by $\pi$ is shorter than the timescale  to change 
$j_1$ by order itself.
Thus, by setting $t_{\rm 1pN}=
t_{\rm LK}j_1$
we find the angular momentum below which LK oscillations are quenched by relativistic precession\footnote{An alternative criterion  can be derived  by requiring  that  at a certain separation,  pericenter precession becomes so strong that the fixed point of the LK problem no longer exists. This leads 
to the following condition for LK oscillations
to be fully quenched by the 1pN pericenter precession \citep{Blaes2002}:
\begin{equation}
\left(\frac{a_2}{a_1}\right)^3 > \frac{3 c^2 m_3 a_1}{4 G (m_1+m_2)^2} \left(\frac{1-e_1^2}{1-e_2^2}\right)^{3/2} 
\label{eqn:suppressed}
\end{equation}
\noindent In the next sections, we use Eqn.~\eqref{eqn:suppressed} to classify systems that will be completely suppressed by pericenter precession, as it is a more conservative criterion than Eqn.~\eqref{eqn:suppressedL}.
} \citep{Antonini2017a}:

\begin{equation}\label{eqn:suppressedL}
j_{\rm GR}= {3G\over \pi  c^2}{(m_1+m_2)^2\over m_3} \left({a_2j_2\over a_1} \right)^3{1 \over a_1 }\ .
\end{equation}

What Equation~(\ref{eqn:suppressedL}) describes is an angular momentum barrier which cannot be passed by systems
that evolve {from} $j_1>j_{\rm GR}$.
Because the pericenter and spin precession terms enter at the same order in the pN expansion, we find that $R= 1$ near the barrier, implying that initially non-adiabatic systems ($R< 1$) cannot 
become adiabatic ($R> 1$) in absence of GW dissipation.  For the population we study in section \ref{sec:popsynth}, roughly 99\% of the BH triples begin their evolution, in the non-adiabatic regime, with 87\% of systems having $R < 0.1$ and 60\% having $R < 0.01$.  At the same time, any triples that may try to evolve into the trans-adiabatic regime is stopped by the angular momentum barrier.  We conclude that any trans-adiabatic (and possibly chaotic) evolution of the SO orientation is suppressed. 
   As such, while we might expect a wide range of SO misalignments from triples in the non-adiabatic regime, especially for cases where the eccentric LK mechanism can produce {orbital} flips of the system \citep[e.g.,][]{Naoz2011}, we do not expect to see truly chaotic evolution of the triple spins as one does for planetary systems.

\subsection{GW Emission and Freezing The Spin Angles}
\label{sec:2.5}

At the peak of a LK oscillation, the eccentricity of the inner binary reaches its maximum, and the energy lost via GW emission can become important for the dynamical evolution of the triple.
We can write the timescale for GW radiation as
\begin{equation}
t_{\rm GW}= a \left | \frac{da}{dt}   \right|^{-1} 
\label{eqn:tgw}
\end{equation}

\noindent where the derivative is given by Equation \eqref{eqn:dadt}.  The value of the angular momentum where GW radiation dominates over the LK dynamics can be derived by setting $t_{\rm GW}=t_{\rm LK} j_1$, which gives:

\begin{equation}
j_{\rm GW}=\left( \frac{170 G m_1 m_2} {3 c^5 m_3} \frac{a_2^3j_2^3}{a_1} \nu ^3\right)^{1/6} \ .
\end{equation}
\noindent For $j_1\leq j_{\rm GW}$ the inner binary effectively decouples from the
third body, and the BBH merges as an isolated system. 

Two situations are relevant here: (i) the 2.5pN terms dominate  the evolution of the binary before 
the 1pN pericenter precession can affect the LK oscillations, and
(ii) the 1pN terms become important before the  2.5pN terms and arrest the evolution of the triple.

In case (i), the inner binary ``plunges'' directly into the regime where GWs dominate its evolution during the maximum eccentricity phase of an
LK oscillation.
After this regime is reached,
the evolution of $e_1$ and $a_1$ are the same
as an isolated binary evolving under GW emission alone; this happens after approximately one LK cycle (starting from $e_1\approx 0$) at the quadrupole level of approximation, or could take several LK cycles at the octupole level. 
We show an example of this type of evolution in the left hand panels of Figure \ref{fig:examples}. The inner binary in this case undergoes extreme ($e_1^{\rm max} > 0.99999$) LK oscillations arising from the strong contribution of the octupole-order terms, including several flips of the inner orbit.  In this strongly non-adiabatic case, the LK oscillations are several orders-of-magnitude faster than the spin precession timescale of the binary, and the spins remain essentially fixed while $\mathbf{j_1}$ varies significantly right up to the merger of the inner binary.  

In case (ii), when the precessional effect becomes important before the
dissipative effects dominate, $e_1$ and the inclination experience damped oscillations, where the angular momentum barrier (Eqn.~\ref{eqn:suppressedL}) suppresses any trans-adiabatic behavior, and $a_1$ and $e_1$ slowly decrease over many LK cycles. 
In the right hand panels of Figure \ref{fig:examples}, we show an example of this type of evolution. Any highly-eccentric oscillations and chaotic spin evolution are dampened by pericenter precession as the inner binary reaches high eccentricities.  The spin vectors can precess onto the invariable plane of the triple \cite[as described in][]{Antonini2017a}, eventually freezing to a constant misalignment with respect to $j_1$ as the inner binary orbit decays due to GW emission to a region where pericenter precession completely suppresses LK oscillations \eqref{eqn:suppressed}.  

The condition for the
transition between regime (ii) and (i) can be  derived by  requiring $j_{\rm GR}<j_{\rm GW}$, which leads to
\begin{eqnarray} \label{rqpn}
a_2j_2< 
\left({\pi c^2 a_1^4 m_3\over G(m_1+m_2)^2}   \right)^{2/5}
 \left( {170\over 3^7} {Gm_1m_2 \nu^3\over c^5 a m_3 }\right)^{1/15} \ .
\end{eqnarray}
  
  For systems that satisfy this condition, the 1pN terms are not important, since  GW radiation will drive 
a fast inspiral of the BBH at $j\approx j_{\rm 2.5pN}$.
Both of these cases have particular observable properties for Advanced LIGO/Virgo which we will explore in Section \ref{sec:observable}.


\section{Triple Population}
\label{sec:pop}

Here we detail the initial conditions considered in this study, and describe our 
method for evolving massive stellar triples from their zero-age main sequence 
birth to the formation of BH triples.  We start with a population of stellar 
triples with the following initial conditions: we draw the primary mass ($m_1$) 
of the inner binary from between $22 M_{\odot}$ and $150 M_{\odot}$ from a 
standard $p(m)dm\propto m^{- \alpha}dm$ distribution, where $\alpha = 2.3$ 
\citep{Kroupa2001}.  The mass of the secondary ($m_2$) is assigned by assuming a 
uniform mass ratio distribution, such that $m_2 / m_1 = U(0,1)$.  The tertiary 
mass ($m_3$) is assigned in a similar fashion ($m_3 / (m_1 + m_2) = U(0,1)$).  
These samples are repeatedly drawn until the initial mass of each star lies 
between $22 M_{\odot}$ to $150 M_{\odot}$.  {The spins for all BHs are assumed to 
be maximal ($\chi = 1$) although we relax this assumption in Section 
\ref{subsec:massandspins}.} 

The orbital properties for the inner and outer binaries are selected from two distinct populations: for the inner binaries, we use recent results on close binaries from \citep{Sana2012}.  The orbital period for the inner binaries are drawn according to $(\log P)^{-0.55}d\log P$ where $P$ is in days and $\log P$ is selected from 0.15 to 5.5.  The eccentricity is drawn from a $p(e)de\propto e^{-0.42}de$ distribution from 0 to 0.9.  For the outer binaries, we assign the semi-major axis from a flat in $\log a$ distribution, while the eccentricity is drawn from a thermal distribution, $p(e)de \propto 2e~de$.  All the angles defining the triple (arguments of pericenter, longitudes of the ascending node, and the inclination) are drawn from isotropic distributions (with the longitude of the ascending node of the outer binary offset from the inner binary by $\pi$ (i.e.~$\Omega_2 = \Omega_1 - \pi$).

To evolve our stellar triples to BH triples, we use a modified form of the Binary Stellar Evolution (BSE) package from \cite{Hurley2002}, including newer prescriptions for wind-driven mass loss, compact object formation, and pulsational-pair instabilities \citep[see details in][]{Rodriguez2016a,Rodriguez2018}.  Each triple is integrated by considering the inner binary and the tertiary as separate stellar systems.  In other words, the binary is evolved using BSE, while the tertiary is evolved using the single stellar evolution (SSE) subset of BSE \citep{Hurley2000}.  This, of course, does not account for the possibilities of mass accretion between the inner binary and the tertiary or any dynamical interaction between the inner and outer binaries (in other words, we do not consider any LK oscillations the triples may experience before they become BH triples).  Such physics, while interesting, is significantly beyond the scope of this paper (though again see \cite{Antonini2017} for a thorough analysis of such triples using the self-consistent method developed in \cite{Toonen2016}).  What we \emph{are} interested in is the change to the orbital components due to the mass loss and BH natal kicks as the elements of the triple evolve towards their final BH states.

To compute this, we track the masses, radii, stellar types, BH natal kicks, and (for the inner binary), the semi-major axes and eccentricities as computed by BSE for every star/binary.  Then at every timestep, we expand the semi-major axis of the outer binary by an increment:

\begin{equation*}
\Delta a_2 = \left(\frac{\Delta M}{m_1 +m_2 + m_3}\right) a_2 .
\end{equation*}

When a BH is formed, we extract the velocity of the natal kick as computed by BSE.  The kick is then applied self-consistently to the orbital elements of the triple \cite[see appendix 1 of][]{Hurley2002}.  Briefly, we assume that each natal kick occurs instantaneously (compared to the orbital timescale).  When the kick occurs, we draw a random orbital phase from the mean anomaly.  The kick is then applied instantaneously to the orbital velocity vector of that component.  We compute the new angular momentum vectors (using the new orbital velocity vector and the same orbital position vector) and a new Laplace-Runge-Lenz vectors, $\mathbf{A}$. This gives us the new orientation of each binary in three-dimensional space, as well as the new semi-major axis and eccentricity, computed via:

\begin{align*}
\mathbf{L}_{\rm{new}}&= M_{\rm new} \mathbf{r}\times \mathbf{v}_{\rm{new}} \\
\mathbf{A}_{\rm{new}} &= \frac{1}{G M_{\rm{new}}}\mathbf{v}_{\rm{new}}\times\mathbf{L}_{\rm{new}}  - \hat{r} 
\end{align*}

\noindent where $M_{\rm{new}}$ is the new mass of the binary post-BH formation, and the new $\mathbf{j}$ and $\mathbf{e}$ vectors are simply $\mathbf{L}_{\rm{new}} / L$ and $\mathbf{A}_{\rm new}$, respectively. The new semi-major axis is given by

\begin{equation*}
a_{\rm{new}} = \frac{2}{r} - \frac{v_{\rm{new}}^{2}}{G M_{\rm{new}}} .
\end{equation*}
%

\noindent If $a_{\rm{new}} < 0$ or $e_{\rm{new}} > 1$, the binary is disrupted.  

Note that we must take care to apply the correct kick to each system. 
{These kicks can be thought of as the combination of two effects: the 
actual change in velocity of one component arising from the asymmetric ejection 
of matereial, 
and the change in the orbital elements from the instantaneous 
loss-of-mass from a single component \citep{Blaauw1961}.  When the inner binary 
undergoes a SN, this mass loss can change the semi-major axis, eccentricity, and 
center-of-mass velocity of the binary.   While most of the changes are natually 
tracked by our above formalism, the change in the center-of-mass velocity of the 
inner binary must be explicitly recorded.  This change is then added to the 
velocity arising from the BH natal kick, and applied as 
$\mathbf{V}_{\rm new}$ to the outer binary}.

In addition to the masses and the evolution of the orbital elements, we implement several additional checks on the survival of our BH triples.  We do not keep any triples which become dynamically unstable at any point during their integration, as a fully chaotic dynamical triple cannot be modeled by the secular evolution considered here (and would very likely result in a collision).  We consider triples to be stable if they satisfy \citep{Mardling2001}:

\begin{equation}
\frac{a_2}{a_1} > \frac{2.8}{1-e_2} \left[\left(\frac{m_1 + m_2 + m_3}{m_1+m_2}\right)\left(\frac{1+e_2}{\sqrt{1-e_2}}\right)\right]^{2/5} .
\label{eqn:marding}
\end{equation}

\noindent We also do not keep any triples which could potentially undergo a collision between their inner and outer components (BSE self-consistently tracks for collisions in the inner binary).  For this, any triple where the pericenter distance of the outer binary can potentially touch the apocenter of the inner binary according to:

\begin{equation*}
a_2 (1 - e_2) - R_3  <  a_1 (1+e_1) + max(R_1,R_2)
\end{equation*}

\noindent at any point during its evolution is discarded (where $R_i$ are the radii of the stars).

Although we will discuss them in Section \ref{sec:popsynth}, we do not dynamically integrate any triple for which the LK mechanism will be strongly suppressed by the 1pN pericenter precession of the inner binary.  For this, we simply exclude any triple for which Equation \eqref{eqn:suppressed} is satisfied once all three objects have evolved to BHs.   Finally, we explicitly exclude from our sample any triples whose inner binary would merge during a Hubble time due to GWs alone.  This is done to limit ourselves to the population of ``useful triples'' \citep[a term taken from][]{Davies2017} those that merge only due to the LK oscillations induced by the tertiary BH.  Although there is likely a small population of merging BBHs whose dynamics may have been altered by a third companion, we do not consider them, in order to maintain a clean separation between the triple-driven BBH mergers studied here and the rate of mergers from common-envelope evolution in galactic fields.

\begin{figure}[]
\centering
\includegraphics[scale=0.9]{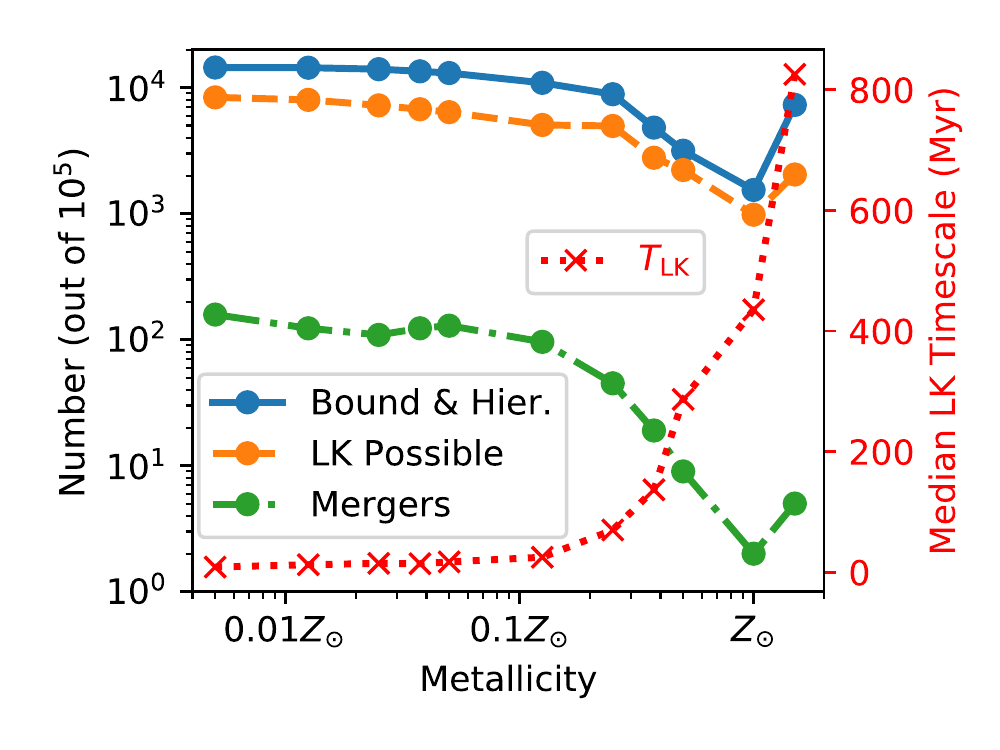}
\caption{The efficiency of stellar triples at producing merging BBH triples as a function of metallicity.  Each metallicity bin started with $10^5$ stellar triples.  Out of those, we show how many triples for which the systems remain bound and hierarchical as they evolve from stars to BHs, the number of those BH triples for which LK oscillations are possible (according to Equation \ref{eqn:suppressed}), and the number of those BH triples.  We also show the median LK timescale (Eqn.~\ref{eqn:tkozai}) in red for all of the LK-possible triples in each metallicity bin.  The $\sim 100$ increase in the number of mergers at low metallicity arises from both the increased number of surviving BH triples and the decreased LK timescale of those triples.}
\label{fig:frac}
\end{figure}

\begin{figure}[]
\centering
\includegraphics[scale=0.98]{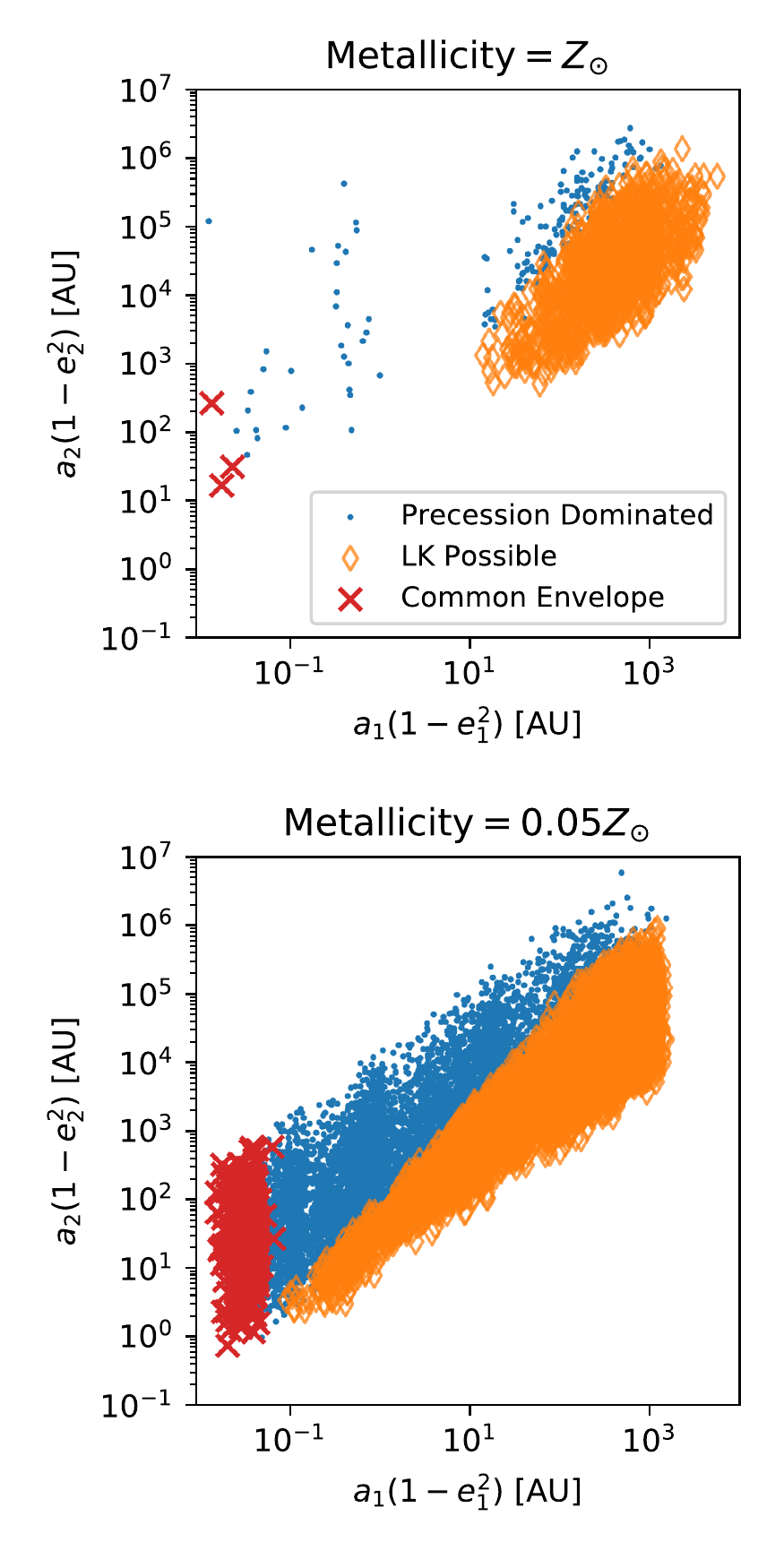}
\caption{The final orbital parameters for the inner and outer binaries of all the bound BH triples at both solar and 5\% solar metallicities.  At solar metallicitiy, the increased mass loss from stellar winds drives both the inner and outer binaries to significantly larger separations (and correspondingly longer LK timescales) compared to low-metallicitiy systems.  We separate the populations into triples that are precession dominated, those for which LK oscillations are possible (Eqn.~\ref{eqn:suppressed}), and those which have undergone a common-envelope phase of evolution.  We find no post-common-envelope systems that have a sufficiently close tertiary companion to undergo LK oscillations.}
\label{fig:single_metal}
\end{figure}

\begin{figure}[]
\centering
\includegraphics[scale=0.85]{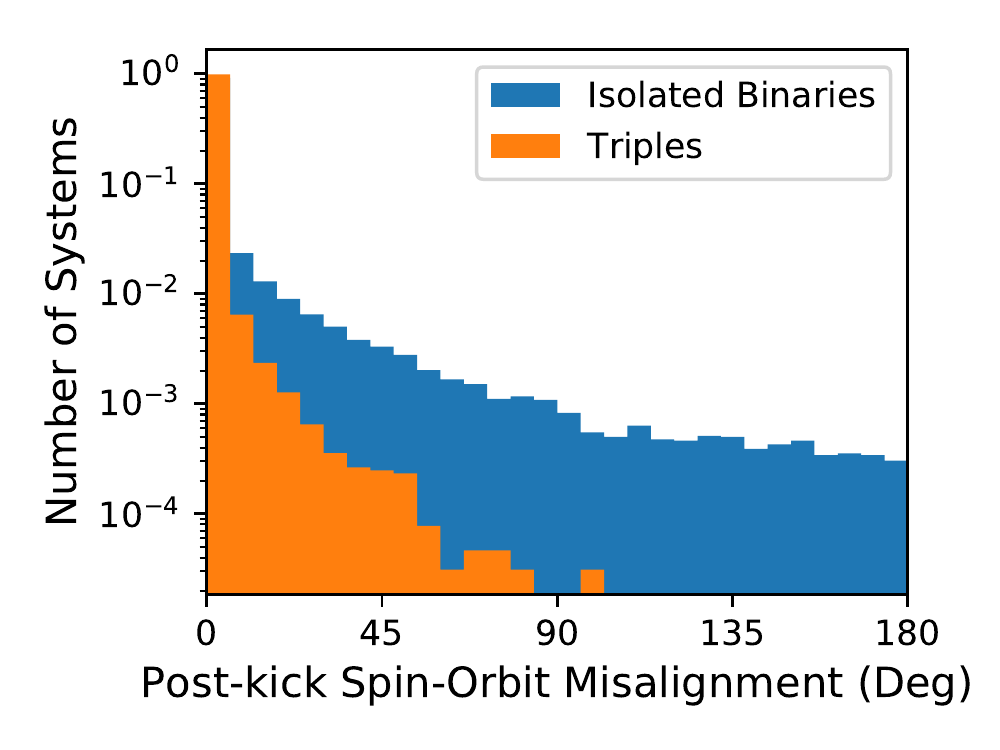}
\caption{The misalignments between the spins of the two BHs and the inner binary angular momentum ($\theta_{{1}}$ and $\theta_{{2}}$) after the triple has evolved to a stable BH triple, assuming the initial stellar spins were aligned with $\mathbf{j_1}$.  We show the final misalignments for all the inner binaries that were evolved (regardless of whether the tertiary remained bound and hierarchical), and for all the triples.  The triple population has a significant preference for spin alignment, since any natal kick capable of significantly torquing $\mathbf{j_1}$ would also likely disrupt the weakly-bound tertiary companion.}
\label{fig:misalign}
\end{figure}

\section{Initial Population of Black Hole Triples}
\label{sec:popsynth}

\subsection{Evolved BH Triples}
\label{subsec:popsynth}

We integrate a population of stellar triples using the formalism and initial conditions described in the previous section.  We consider 11 different stellar metallicities (1.5$Z_{\odot}$, $Z_{\odot}$, 0.5$Z_{\odot}$, 0.375$Z_{\odot}$, 0.25$Z_{\odot}$,
0.125$Z_{\odot}$,  0.05$Z_{\odot}$, 0.0375$Z_{\odot}$, 0.025$Z_{\odot}$,
0.0125$Z_{\odot}$, and 0.005$Z_{\odot}$), and integrate $10^5$ different triples for each metallicitiy bin, for a total of $1.1\times 10^6$ stellar triples.

Of immediate interest is the number of stellar triples which survive to become stable, hierarchical BH triples.  Of our $1.1\times 10^6$ triples, approximately $10\%$ evolve from zero-age main sequence triples to hierarchically-stable (Eqn.~\ref{eqn:marding}) BH triples without being disrupted due to mass loss, natal kicks, or undergoing stellar collisions.  Of those, roughly half ($5\%$ of the initial population) form triples where the LK timescale is less than the precession timescale of the inner binary, and have the potential to undergo LK oscillations (Eqn.~\ref{eqn:suppressed}).  However, these results are highly dependent on the metallicitiy of the system.  

In Figure \ref{fig:frac}, we show the number of systems in each metallicitiy bin 
which survive their evolution from stellar triple to BH triple to LK-induced 
merger (which we will explore in the next section).  As the metallicitiy is 
increased, the number of systems which survive their evolution decreases 
dramatically, with nearly an order-of-magnitude fewer systems remaining as bound 
and hierarchical triples at solar metallicitiy than at lower ($\sim 
0.01Z_{\odot}$) metallicities.  This is entirely due to the mass loss and natal 
kicks experienced by massive stars at different stellar metallicities.  Massive 
stars with high metallicities lose significant amounts of their mass during 
their evolution, largely due to radiation pressure in higher-opacity envelopes and line-driven winds \cite[e.g.,][]{Vink2001}.  This mass loss expands the bound systems (such as binaries and triples), making each 
system more susceptible to disruption during stellar collapse.  Furthermore, 
high-metallicity stars lose more mass, producing lower-mass cores and correspondingly
lower-mass BHs.  These systems are conjectured to experience large natal kicks 
during supernova (SN), adding significant velocity kicks to the system \citep[e.g.,][]{Fryer2012,Repetto2015}, while many of the massive BHs that form at lower metallicities are 
expected to form via direct collapse, experiencing little to no kick 
\citep{Fryer2001,Belczynski2016}.  The effect of metallicitiy is two-fold: high-metallicity systems lose more mass during their evolution, significantly expanding their orbits, where the stronger natal kicks associated with these lower-mass BHs can more easily disrupt the outer orbits.

While the survival of BH triples increases by nearly a factor of 10 at low $Z$ versus $Z_{\odot}$, the number of LK-induced mergers 
increases by almost a factor of 100.  This additional increase is due 
to the decreased mass loss at lower metallicities, which reduces the expansion of both the inner and outer semi-major axes during 
the evolution of the triples.  These tighter triples have significantly shorter
LK timescales than triples at high metallicities.  In Figure \ref{fig:frac} (red axis) we show the median 
LK timescale \eqref{eqn:tkozai} for the collection of bound and 
hierarchical triples in each metallicity bin.  The typical LK timescale of 
the lowest metallicitiy triples is $\sim 100$ times shorter than those at solar metallicities.  These triples will have many opportunities to 
experience highly-eccentric oscillations (thousands per Hubble time) that may 
induce a merger, while high metallicity systems may undergo only a few to tens of 
oscillations within the age of the Universe.

In Figure \ref{fig:single_metal}, we show the orbital parameters of the inner 
and outer binaries for those bound and hierarchical triples in our $Z_{\odot}$ and 
$0.05Z_{\odot}$ models. While the minimum value of $a_1(1-e_1^2)$ at solar metallicity for LK-driven 
mergers is $\sim 10$ AU, the low-metallicitiy triples span the allowed 
range of inner orbits from 0.1 to $10^3 a_1(1-e_1^2)$.  Below 0.1, we find that 
all BH triples are suppressed from undergoing LK oscillations by the pericenter 
precession of the inner binary.  We note that this includes all systems for 
which the inner binary has undergone a common envelope phase of evolution.  Although we find many post-common-envelope systems in our $1.1\times10^6$ 
triples, they all inhabit the region of parameter space  
(Eqn.~\ref{eqn:suppressed}) where pericenter 
precession quenches any possibility of LK oscillations.

\subsection{Initial BH spin-misalignment}
\label{subsec:misalignment}

Although the formalism presented in Section \ref{sec:eom} allows us to track the evolution 
of the spin vectors for the hierarchical three-body problem, the initial amount 
of misalignment between the BH spins and the inner binary angular 
momentum must be treated carefully.  While we assume that the initial 
stellar spins are aligned with $\mathbf{j_1}$, it is well 
known that the BH kicks can significantly misalign the orbital and spin 
angular momenta \cite[e.g.][]{Kalogera2000}, since any instantaneous kick to one of the binary components will change the direction of the orbital angular momentum.  Fortunately, the formalism presented in 
\cite{Hurley2002} and employed here makes it trivial to track the change in 
orientation of $\mathbf{j_1}$ through the two SN of the inner binary (the outer 
binary kick cannot torque the inner binary).

In Figure \ref{fig:misalign}, we show the post-kick misalignments of the BH 
spins with $\mathbf{j_1}$, assuming that
SN kicks are emitted isotropically in 
direction from the surface of the exploding stars, and that neither mass transfer nor tidal 
torques can realign either the stellar or BH spin with the orbit after the natal kick.  These are 
both highly conservative assumptions allowing us to explore the maximum allowed 
post-SN misalignment \citep[see][]{Rodriguez2016c}.  We find that the vast 
majority of post-kick misalignment are very small, with $97\%$ of systems having 
misalignments less than 0.1 degrees, and $99\%$ of systems having misalignments 
less than 6 degrees.  On the other hand, the misalignments of the isolated binaries 
themselves (ignoring whether the third BH remains bound) can be significantly larger, with $\sim 8\%$ 
of binaries having misalignments greater than 6 degrees 
\citep[in agreement with][]{Rodriguez2016c}.  This increased preference for aligned triples arises 
from the difficulty of keeping the outer companion bound to the inner binary post-SN.  While the 
semi-major axes for the inner binaries are sufficiently small for the binary to 
survive the kick, the outer orbits are so wide that the SN kicks are 
frequently several times larger than the typical orbital velocities for the 
outer orbits (usually a few $\rm{km}/\rm{s}$).  Since smaller kicks produce 
smaller spin-orbit misalignments, the requirement that the tertiary companion 
remains bound significantly limits the possible range of initial 
misalignments.

Because of that, and the associated difficulties of following spin realignment 
through mass transfer and tidal torques, we will assume that our 
BH triple systems begin with their BH spins aligned with $j_1$ (although we test the implications of this assumption in Figure \ref{fig:chi_align}).

\begin{figure}[]
\centering
\includegraphics[scale=0.85]{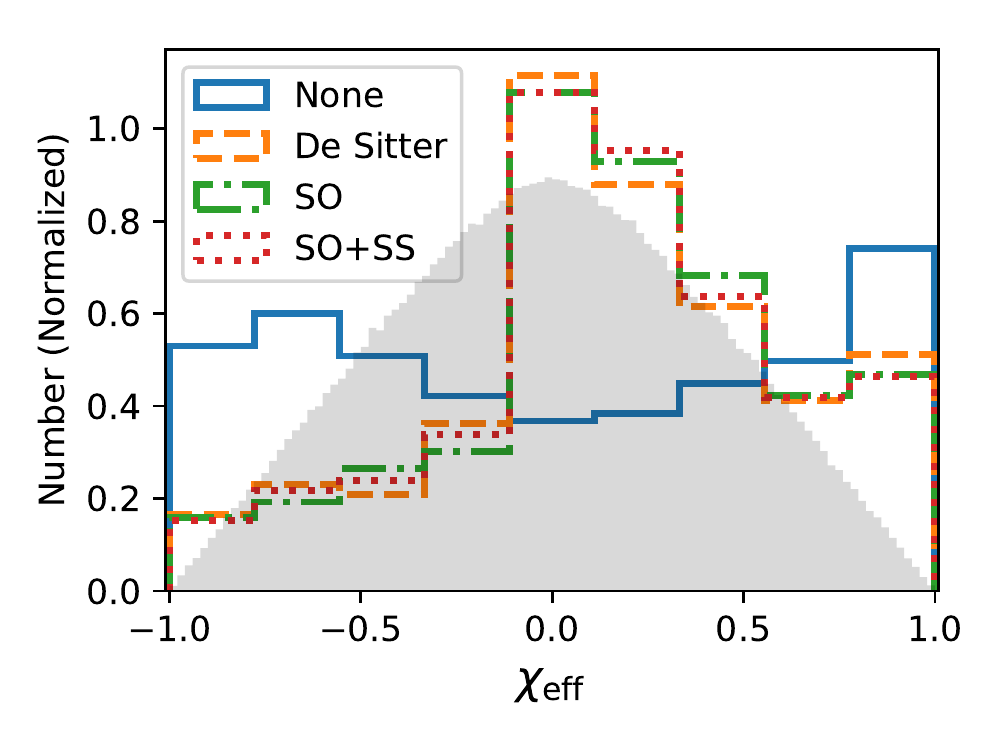}
\caption{The distribution of effective spins for merging BBHs.  We show the 
distribution for triples with varying levels of spin physics.  We consider 
triples with no spin effects, only 1pN precession of the spins 
(Eqn.~\ref{eqn:dsdtso}), all SO terms (Eqns.~\ref{eqn:djdtso}-\ref{eqn:dsdtso}), 
and all SO and SS terms (Eqns.~\ref{eqn:djdtso}-\ref{eqn:dsdtss}).  Considering 
only the 1pN ``de Sitter'' precession of the spins is clearly sufficient for 
most cases, since the higher-order SO and SS terms to not play a significant 
role until immediately before merger.  The filled gray histogram shows 
$\chi_{\rm eff}$ if the spins were completely isotropic ({but using the 
mass distribution from our population synthesis}).}
\label{fig:spin_terms}
\end{figure}

\begin{figure}[tb]
\centering
\includegraphics[scale=0.85]{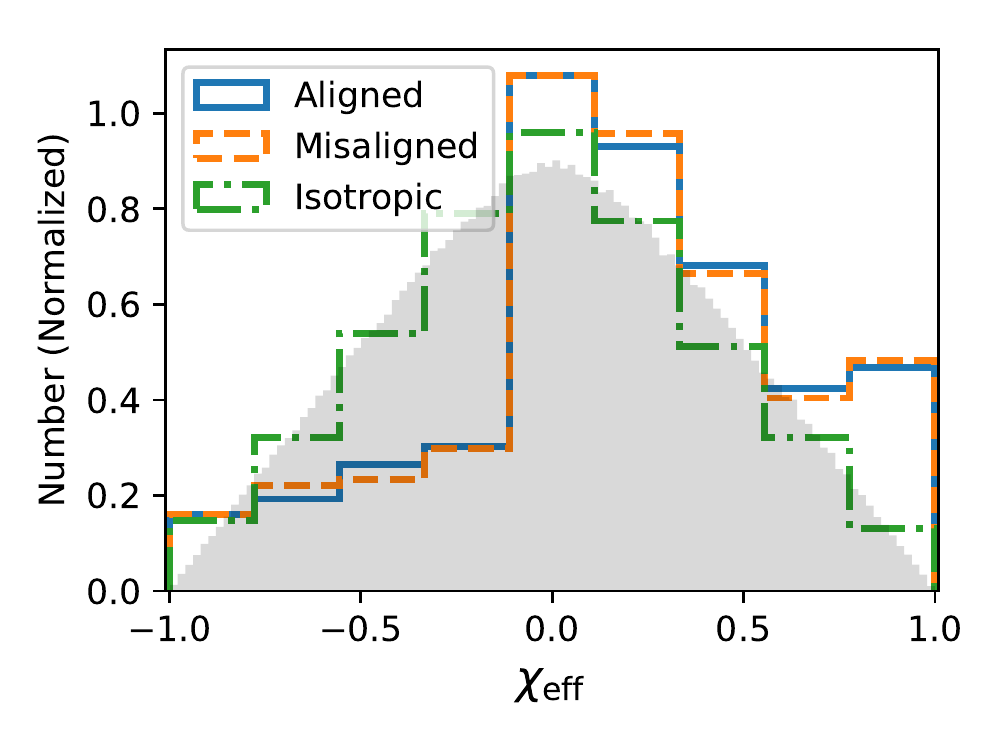}
\caption{Similar to Figure \ref{fig:spin_terms} (with all spin terms), but considering different initial spin alignments for the inner BBHs.  We show the final $\chi_{\rm eff}$ distributions for BBHs which being perfectly aligned with $\mathbf{j_1}$, those that begin with some misalignment based on BH natal kicks (Figure \ref{fig:misalign}), and a completely isotropic initial distribution.  The aligned and misaligned distributions are virtually identical, largely because the initial misalignments are extremely small.  The initially isotropic distribution yields a final distribution that is also isotropic. }
\label{fig:chi_align}
\end{figure}

\section{Spin Distributions of Useful Triples}
\label{sec:spin}


We now turn to understanding the spin 
distributions of merging triple systems formed from stellar triples.  As stated previously, we are interested in the distribution of ``useful'' triples, which we define to be those mergers whose inner binaries would not have merged in a Hubble time as an isolated system.  We focus 
mainly on the distributions of $\chi_{\rm{eff}}$ (Eqn.~\ref{eqn:chieff}), as this is 
the spin parameter most easily measured by Advanced LIGO/Virgo.  The first 
question that naturally arises is whether the 
inclusion of higher-order pN corrections (the SO and SS terms) 
have a significant influence on the final measurable values of 
$\chi_{\rm{eff}}$.  It was claimed in \cite{Liu2017,Antonini2017,Liu2018} that the 
lowest-order precession of the spin vectors about $\mathbf{j_1}$ (also known as 
geodedic or de Sitter precession) would dominate the spin dynamics of the system,  
with the higher-order terms (such as the Lense-Thirring/SO coupling or 
the SS coupling) would not significantly effect the dynamical evolution.

In Figure \ref{fig:spin_terms}, we show the $\chi_{\rm{eff}}$ distributions of our 
useful triples, and how they vary depending on the sophistication of the 
SO physics considered.  As expected \citep{Antonini2017}, 
there is a significant difference 
between the triples integrated with no spin terms and those integrated by 
considering de Sitter precession (which we simulate by including all SO 
terms, but setting $\chi_{1} = \chi_{2}=0$, allowing the spin vectors to precess 
but not couple to the orbits). 

Figure \ref{fig:spin_terms} also makes clear that the inclusion of the full 
SO and SS terms do not play a significant role in the final 
distribution of the measurable spin terms.  This is fully consistent with the discussion in Section \ref{sec:2.5}: any triple for which the separation during LK oscillations would get small enough for SO or SS effects to become relevant will immediately decouple from the tertiary and merge due to the GW emission.  This ``case (i)'' type of merger is illustrated in the left hand panel of Figure \ref{fig:examples}.  On the other hand, for any case where the LK oscillations remain relevant during the inspiral, the 1pN pericenter precession will suppress any higher-order pN effects until the binary effectively decouples from the third companion (``case (ii), or the right panels of Figure \ref{fig:examples}).  In other words, there exists no regime in which the SO or SS terms are relevant for the distributions of $\chi_{\rm eff}$.

We have also assumed that the initial spin distributions of the BHs are aligned 
with $\mathbf{j_1}$ at the beginning of the LK evolution of the triple.  We 
showed in Section \ref{subsec:misalignment} that, due to the natal kicks, the vast majority of bound and hierarchical BH triples have spins 
initially aligned with $\mathbf{j_1}$ (assuming the spins of the stars were initially 
aligned with $\mathbf{j_1}$).  In Figure \ref{fig:chi_align}, we show $\chi_{\rm{eff}}$ for the same 
population integrated with all spin terms, but with different initial 
orientations for the spins.  As expected, the initial misalignments for the 
stellar triples make little difference in the final $\chi_{\rm{eff}}$ 
distributions.  However, we also find that, if the distribution of initial spins 
is isotropic \citep[as would be expected for dynamically-assembled 
triples, e.g.,][]{Antonini2016}, then the final distribution of spins is 
also isotropic.  This is to be expected, since it is well known that an 
isotropic distribution of spins will remain isotropic during 
inspirals of isolated binaries \citep{Schnittman2004}, and there is no 
reason to expect that differential precession of the two spin vectors by de 
Sitter precession (which dominates the spin evolution during LK 
oscillations) to create a preferred direction from an isotropic distribution.

\section{Gravitational-wave Observables}
\label{sec:observable}

\subsection{Masses and Spins}
\label{subsec:massandspins}

\begin{figure}[bt]
\centering
\includegraphics[scale=0.9]{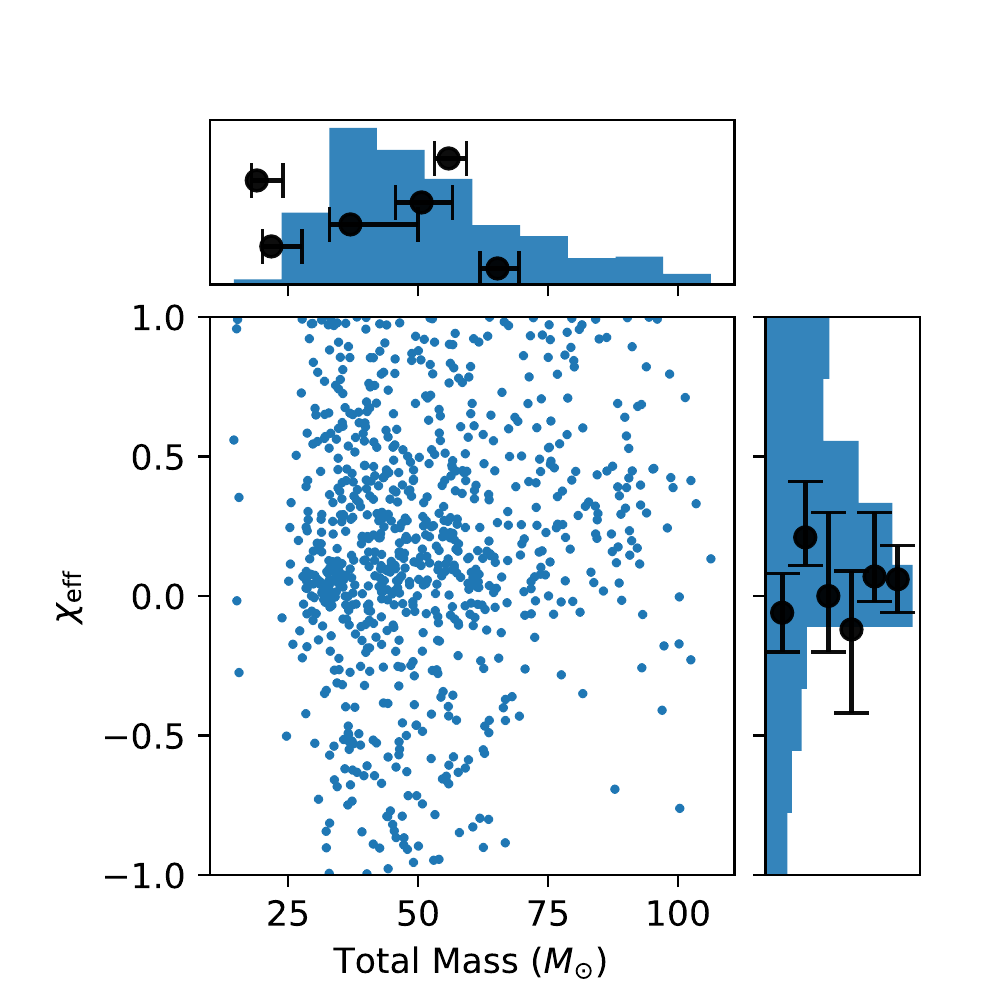}
\caption{Joint total mass ($m_1 + m_2$) and effective spin distributions for merging BBHs.  We also show each of the measured posteriors for $M_{\rm tot}$ and $\chi_{\rm eff}$ for each of the 5 BBH mergers (and 1 BBH merger candidate) reported by LIGO/Virgo so far. }
\label{fig:mvchi}
\end{figure}

The first obvious observable parameters that can be explored by the current generation of gravitational-wave detectors are the masses and the spins of the merging BHs.  As mentioned previously, the spin parameter most easily measured by Advanced LIGO/Virgo is the effective spin parameter, $\chi_{\rm eff}$.  One can immediately ask whether there is any immediate correlation between the masses and the effective spins of our merging triples.  

In Figure \ref{fig:mvchi}, we show the 2D distributions of the total mass versus the effective spin.  There is no strong correlation between the effective spin and the total mass for the population of triple-driven mergers surveyed here.  This is to be expected, since the LK timescale \eqref{eqn:tkozai} is much more strongly dependent on the angular momentum of the outer binary than the masses of the triple components.  While this may not hold for more massive systems (such as a stellar mass BBH in orbit around a super-massive BH), for the cases considered here there is no significant correlation.  We do note that this model naturally explains all of the heavy BBHs observed to date (those with total masses $\gtrsim 40M_{\odot}$) along with their correspondingly low effective spins.

\begin{figure}[tbh]
\centering
\includegraphics[scale=0.8]{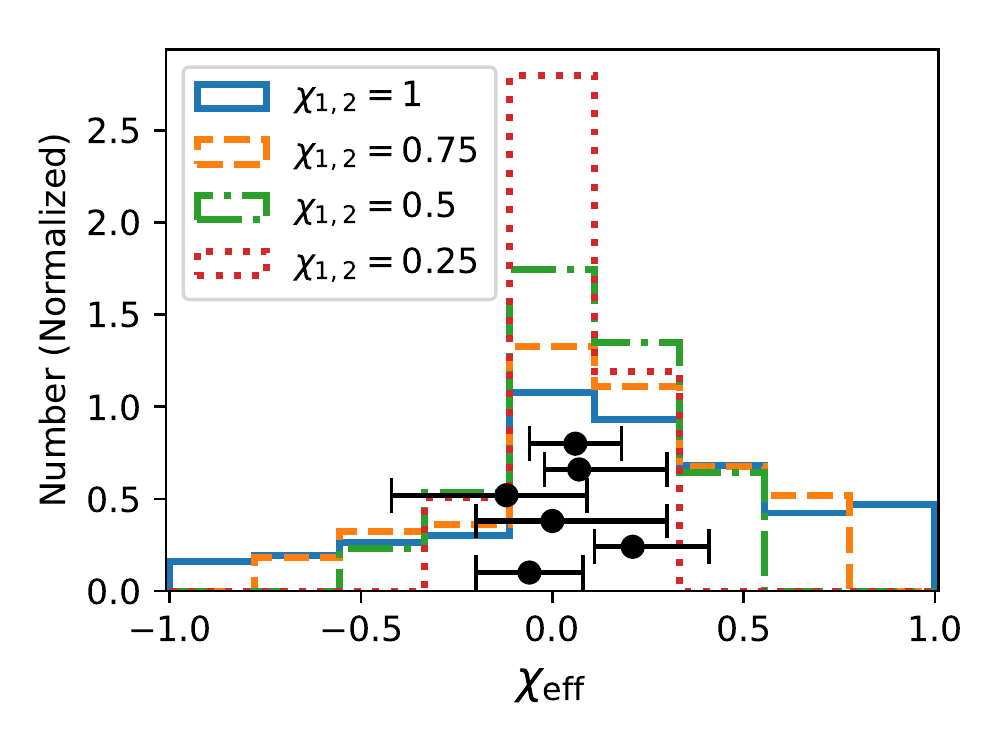}
\caption{The same as the top panel of Figure \ref{fig:mvchi}, but now showing the $\chi_{\rm eff}$ distribution as a function of different spin magnitudes, calculated using the same final spin angles from the previous section.  Although this ignores the reduced SO coupling from the lower spin magnitudes, we have shown (Figure \ref{fig:spin_terms}) that the SO and SS couplings do not influence the distributions of $\chi_{\rm eff}$, and the 1pN precession of the spins is independent of the spin magnitudes.  The black points and their error bars show the 6 $\chi_{\rm eff}$ measurements reported by LIGO/Virgo so far.}
\label{fig:chi_low}
\end{figure}

Throughout this analysis, we have assumed that the spins of the BBHs are maximal.  This was done for simplicity, but we can also consider the effect of spin magnitudes on our predictions for $\chi_{\rm{eff}}$.  Because we have shown that the SO and SS terms have a negligable effect on the spin evolution (while the precession of the spins about $\mathbf{j}_1$ is independent of the spin magnitudes), for $\chi_{\rm{eff}}$ in these systems, we can to good approximation simply replace the spin magnitudes in our integrated $\chi_{\rm{eff}}$ distributions, to determine what $\chi_{\rm{eff}}$ would have been with lower spins.  We show these values in Figure \ref{fig:chi_low}.  As the spin magnitudes are decreased, the distribution of $\chi_{\rm{eff}}$ converges to zero, as would be expected for systems with low spins lying in the plane. 

While we have focused on the effective spin of the BBHs, in reality the spin information is much more complex.  It has been suggested \citep{Gerosa2013,Gerosab,Trifiro} that a complete measurement of the BBH spin angles will allow Advanced LIGO/Virgo to discriminate not only between different formation channels for BBHs, but to measure differences in the binary stellar physics producing BBH mergers from isolated field binaries.  

\begin{figure}[tbh]
\centering
\includegraphics[scale=0.85]{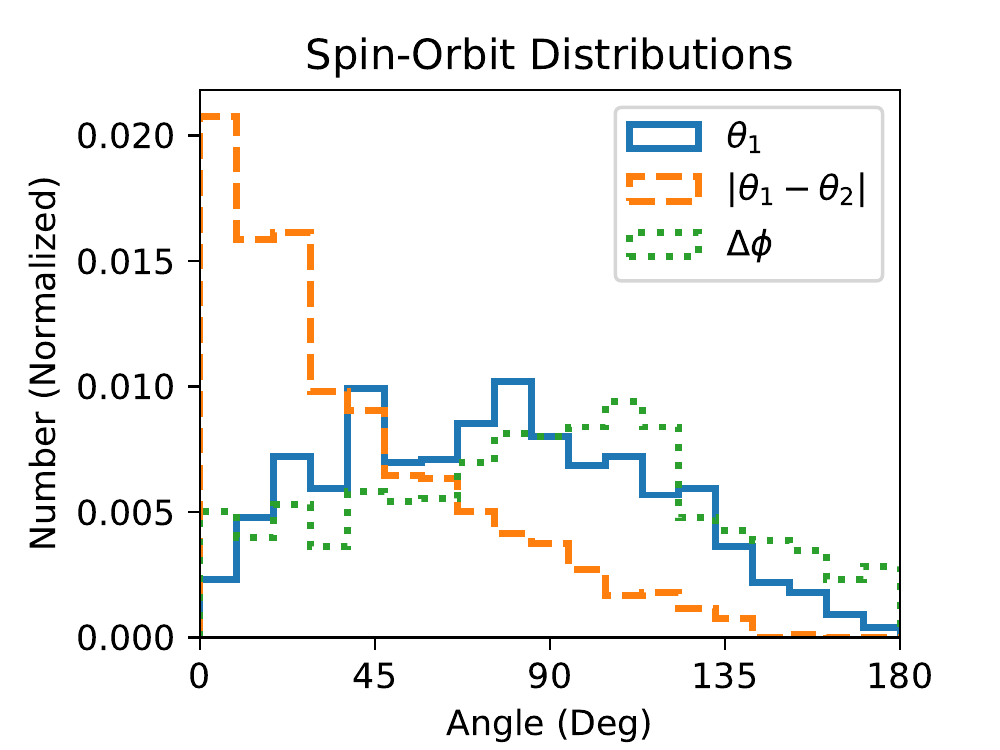}
\caption{Final distributions of the individual spin vectors with respect to $\mathbf{j_1}$.  We show the absolute misalignment between $\mathbf{S_1}$ and $\mathbf{j_1}$, $\theta_1$, the relative misalignment angles between $\mathbf{S_1}$ and $\mathbf{S_2}$ along the azimuth ($\lvert\theta_1  -\theta_2\rvert$) and, the angle between the two spins in the orbital plane of the binary (perpendicular to $\mathbf{j_1}$, $\Delta \phi$).}
\label{fig:sodist}
\end{figure}

In Figure \ref{fig:sodist}, we show the distributions of $\theta_1$, $|\theta_1 - \theta_2|$, and $\Delta \phi$.  Both $\theta_1$ and $\Delta \phi$ 
have broad distributions.
The former arises from spin precession described in the previous sections, while the later is a natural feature of randomly distribution vectors in the plane \cite[see e.g.,][their Figure 2]{Gerosa2013}.  At the same time, we note that our distribution of  $|\theta_1 - \theta_2|$ is somewhat broader than the one presented there.  This suggests that sufficient observations of the individual spin angles can be used not only to better understand binary stellar evolution, but to discriminate between binary and triple stellar evolution of BBH mergers \citep[e.g.,][]{Trifiro}.

\subsection{Eccentricity and Spin}
\label{subsec:eccspin}

The eccentricity distribution of BBHs from LK-induced mergers is one of the distinct observables that can be identified by GW detectors.  While BBHs from isolated field binaries and BBHs ejected from globular clusters will reach the LIGO/Virgo band with very low orbital eccentricities \citep[$\lesssim 10^{-3}$, e.g.][]{Breivik2016}, the presense of the third companion can induce highly eccentric mergers which can maintain  eccentricisties as high as $e\sim 0.1$ up to a GW frequency of 10Hz \citep[assuming the secular approximation to be valid, e.g.][]{Antonini2014}.  How these values correlate with the spins can be a significant discriminant between BBH formation channels.

\begin{figure}[]
\centering
\includegraphics[scale=0.95]{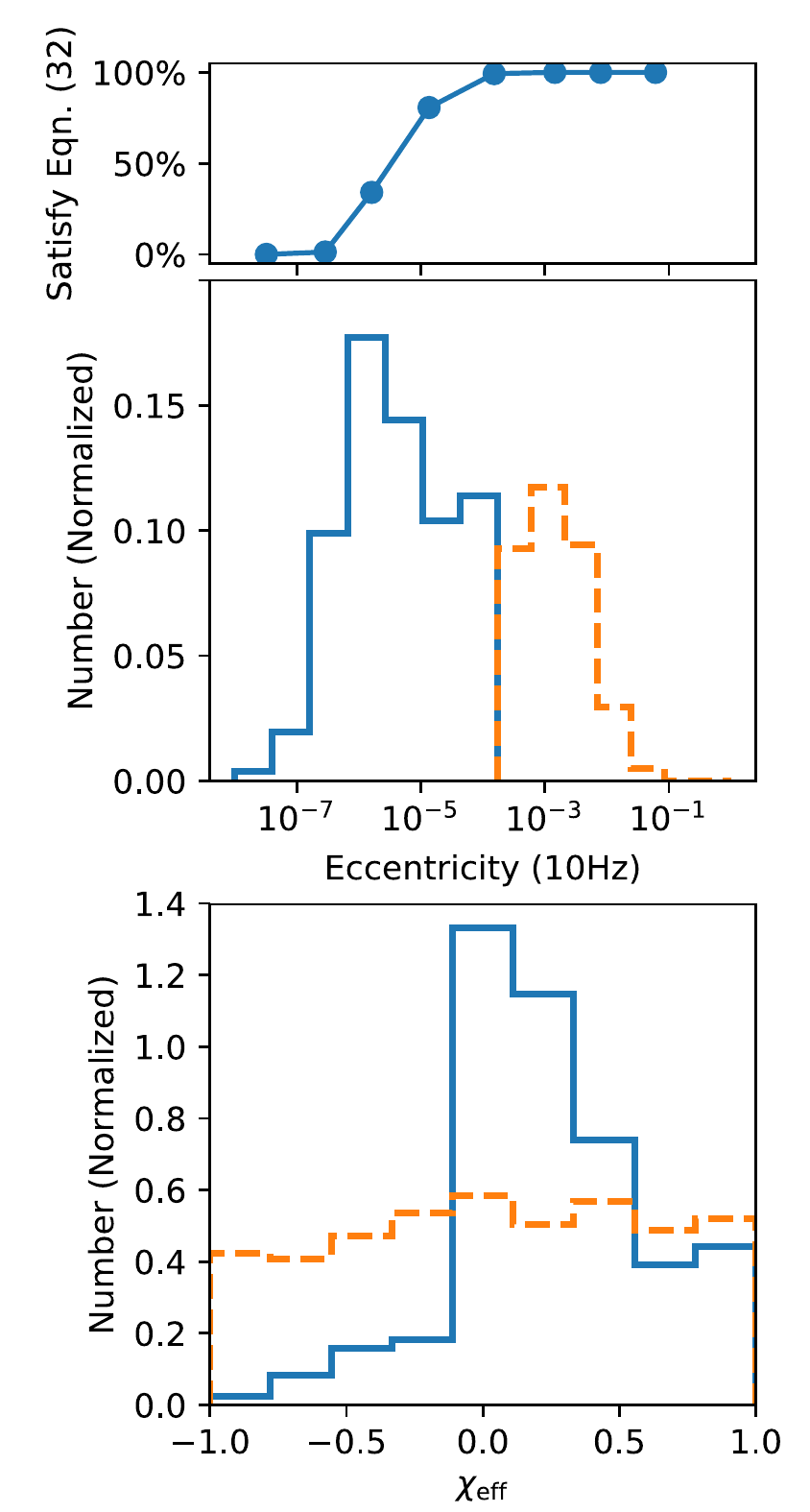}
\caption{The final eccentricity (at a GW frequency of 10Hz) and $\chi_{\rm{eff}}$ distributions from merging BBHs.  On the top, we show the eccentricity distributions above and below $2\times10^{-4}$ (in solid blue and dashed orange), while the insert along the top shows what fraction of the triples at each eccentricity initially satisfies Equation \eqref{rqpn}.  On the bottom, we show the $\chi_{\rm eff}$ distributions corresponding to those same eccentricity bins.  For high eccentricity systems, the distribution is nearly flat, while low eccentricity distributions clearly show the distinct $\chi_{\rm eff}$ distribution shown in the text.  An example from each of these distributions is shown in Figure \ref{fig:examples}, the left and right panels respectively.}
\label{fig:eccchi}
\end{figure}

In Figure \ref{fig:eccchi}, we show the 1D distributions for the eccentricity at a GW frequency of 10Hz (the lower-bound of the LIGO band) and their corresponding $\chi_{\rm eff}$ distributions.  There is a clear break at $e_{\rm 10Hz}\sim 2\times 10^{-4}$, where binaries with higher eccentricities have a nearly flat distribution in $\chi_{\rm eff}$ from -1 to 1, while binaries with lower eccentricities recover the $\chi_{\rm{eff}}\sim 0$ peak described in the previous sections. 
This behavior is discussed at length in Section \ref{sec:2.5}, and is well illustrated in Figure \ref{fig:examples}.  In the case (i) example (left panels), the strong octupole terms from the interaction Hamiltonian drive the eccentricities to very large values ($e_1 > 0.99999$) such that GW emission drives the binary to merge before the 1pN terms can significantly dampen the LK oscillations or cause the spins to precess.  At the same time, the inner binary can flip its angular momentum several times, while the spin-orbit angles vary wildly.  When this binary merges (essentially in a single highly-eccentric oscillation), the spins are still aligned with each other, while the spin-orbit 
orientation is drawn from a nearly random distribution
from 0 to 180$^{\circ}$. This type of evolution 
results  in the uniform $\chi_{\rm eff}$ distribution of the higher eccentricity systems displayed in Figure \ref{fig:eccchi}. On the other hand, the smoother, case (ii) evolution in the right panels of Figure \ref{fig:examples} does not experience large eccentricity/inclination oscillations, since any highly-eccentric oscillations are arrested by the angular momentum barrier (Eqn.~\ref{rqpn}).  This significantly longer evolution allows the spins to experience significant precession 
 producing a $\chi_{\rm eff}$ near 0.5 \citep{Antonini2017a}, while the angular momentum barrier keeps the maximum eccentricity at lower values, yielding a lower eccentricity at merger.  We show the fraction of systems which obey Equation \eqref{rqpn} as a function of eccentricity in the top panel of Figure \ref{fig:eccchi}.
 
 We do note that, by restricting ourselves to the secular equations of motion, we have explicitly ignored the highly-eccentric mergers that can occur during LK oscillations when the secular approximation breaks down \citep[e.g.,][]{Antonini2014}.  Since the breakdown occurs in regimes where very high eccentricities allow GW emission to change the triple on an orbital timescale, these systems (with $e_{10\rm{Hz}}\sim 1$) would likely show a similarly flat distribution in $\chi_{\rm eff}$.

\subsection{Merger Rate}
\label{subsec:rate}

To probe the contribution from this channel on the population of BBH binaries detected by Advanced LIGO/Virgo, we place our models of BH triples into a cosmological context.  We begin by assuming that the formation of stellar BH triples will follow the cosmological star-formation rate (SFR) of the universe.  We use the SFR as a function of redshift from \cite{Belczynski2016}, based on significant multi-wavelength observations \citep[see e.g.,][]{Madau2014}

\begin{equation}
\rm{SFR}(z) = 0.015 \frac{(1+z)^{2.7}}{1 + ((1+z)/2.9)^{5.6}} M_{\odot}\rm{Mpc}^{-3} \rm{yr}^{-1}
\label{eqn:sfr} .
\end{equation}

\noindent Because we have shown that the contribution from low-metallicity star formation is the dominant contribution to the BH triple channel (e.g.~Figure \ref{fig:frac}), we consider only the SFR for stars with $Z < 0.25 Z_{\odot}$.  This is done by computing the cumulative fraction of star formation, using the chemical enrichment model of \cite{Belczynski2016}.  In this model, the mean metallicitiy $Z$ at a given redshift $z$ is given by

\begin{align}
\log_{10} \left<Z(z)\right> = 0.5+ \log_{10} &\Big[ \frac{y (1-R)}{\rho_b}\nonumber \\ &\times \int^{20}_{z} \frac{97.8\times10^{10} \rm{SFR(z')}}{H_0 E(z')(1+z')}dz \Big]
\label{eqn:enrichment}
\end{align}

\noindent where $R = 0.27$ is the fraction of mass from each generation of stars that is returned to the interstellar medium, $y = 0.019$ is the mass fraction of new metals created in each generation of stars, $\rho = 2.77\times10^{11} \Omega_b h_0^2 M_{\odot} \rm{Mpc}^{-3}$ is the baryon density with $\Omega_b = 0.045$ and $h_0 \equiv H_0/100$, and $E(z) \equiv \sqrt{\Omega_M(1+z)^3 + \Omega_k (1+z)^2 + \Omega_\Lambda}$.  

\begin{figure}[]
\centering
\includegraphics[scale=0.88]{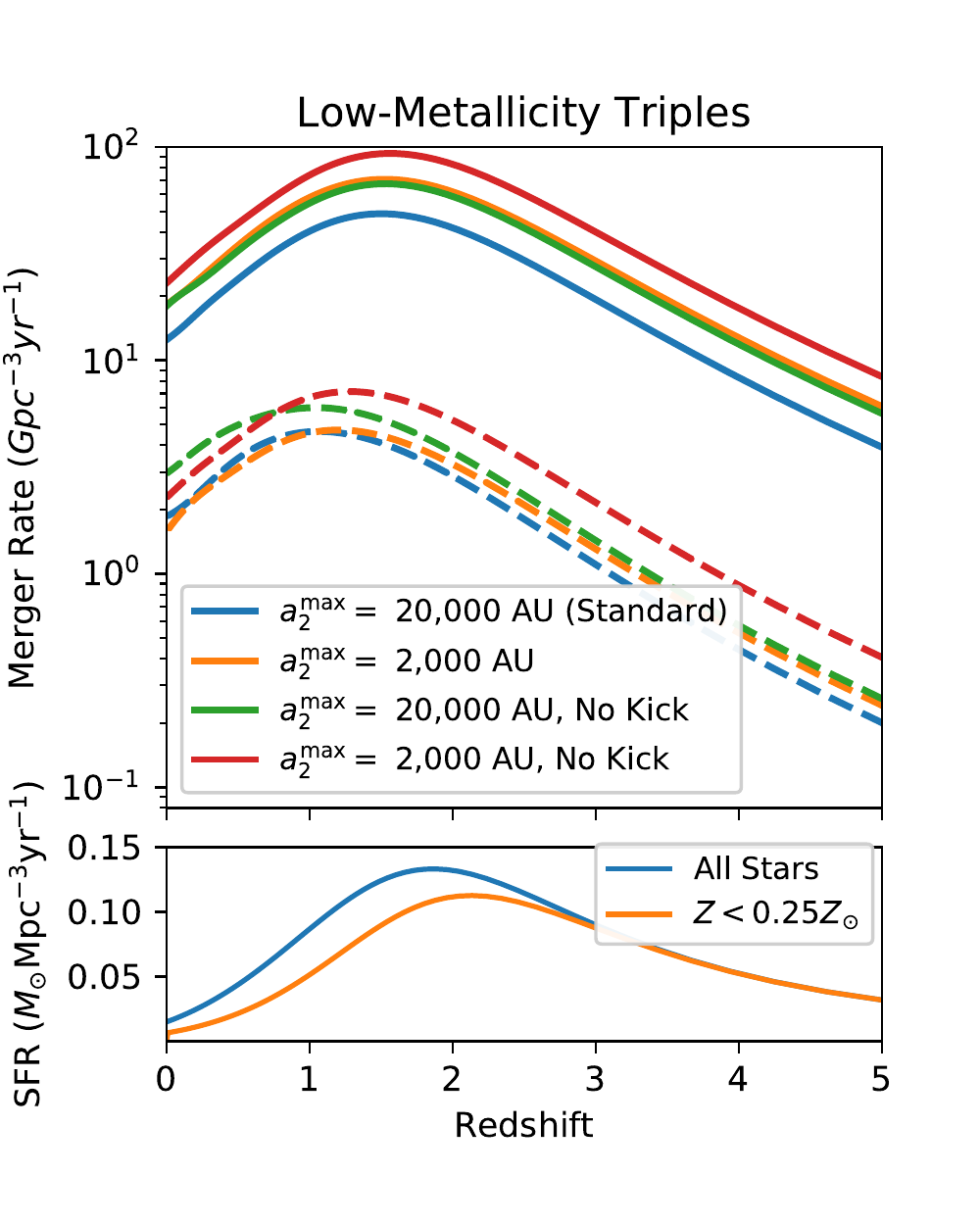}
\caption{The merger rate of low-metallicitiy BH triples as a function of cosmological redshift.  In the top panel, we show the four variant triple BH populations described in the main text (Section \ref{subsec:rate}).  We bracket the uncertainties in our triple stellar evolution by assuming that i) no triples merge or interact during their main sequence evolution (solid lines), or ii) any stellar triple that can undergo LK oscillations with a LK timescale $< 3$Myr merges during its main-sequence evolution (dashed lines) and is excluded from our rate estimate.   The pessimistic criterion preferentially selects triples with longer delay times between formation and merger, which in turn pushes the peak of the pessimistic estimate to lower redshifts. In the bottom panel, we show our assumed SFR, and the SFR rate when restricted to metallicities below 0.25 $Z_{\odot}$ (Equations \ref{eqn:sfr}-\ref{eqn:sfrlow}).}
\label{fig:rate}
\end{figure}

In \cite{Belczynski2016}, they assumed that the distribution of metallicitiy at a given redshift followed a log-normal distribution, with mean given by \eqref{eqn:enrichment} and a standard deviation of 0.5 dex \citep[based on measurements from ][]{Dvorkin2015}.  Since we are only interested in metallicitiy below $0.25 Z_{\odot}$, and since we are dealing with small number statistics, we convolve the SFR from \eqref{eqn:sfr} with the cumulative distribution of metallicities:

\begin{align}
\rm{SFR}_{\rm{Z < Z_{\rm{low}}}}(z) &= \rm{SFR}(z) \\ \times &\left(1 + \rm{erf} \left[\frac{\log_{10} (Z_{\rm{low}}) - \log_{10} \left<Z(z)\right>}{0.5 \sqrt{2}}\right]\right) \nonumber .
\label{eqn:sfrlow}
\end{align}

For this estimate, we restrict ourselves to models for which $Z < 0.25 Z_{\odot}$ (see the bottom panel of Figure \ref{fig:rate}).  We then assume that the rate can be expressed as 

\begin{align}
R(z) = F_{\rm{merge}}& F_{\rm triple}  F_{M > 22M_{\odot}}\left<M_{\star}\right> \noindent \\
\times &\int_{-\infty}^{\infty} t_{\rm{delay}}(t(z) - t')\rm{SFR}_{\rm{Z < Z_{\rm{low}}}}(z(t')) dt' \nonumber
\end{align}

\noindent  where $F_{\rm{merge}}$ is the fraction of triples which merge (e.g. Figure \ref{fig:frac}), $F_{\rm triple}$ is the fraction of massive binaries with a tertiary companion \citep[taken to be 60\%, following][their Figure 16]{Sana2014},  $F_{M > 22M_{\odot}}$ is the fraction of triples with all three components having masses above $22M_{\odot}$ (assuming $m_1$ is drawn from the IMF, while $m_2$ and $m_3$ are drawn from distributions uniform in the mass ratio), $\left<M_{\star}\right>$ is the mean mass of a star from the IMF (both taken from \cite{Kroupa2001}, running from $0.01M_{\odot}$ to $150M_{\odot}$), and $t_{\rm delay}$ is the distribution of delay times between triple formation and merger.  The integral is the convolution of the delay time distribution over the low-metallicity SFR.  The factors of $z(t)$ and $t(z)$, giving the redshift at a given lookback time and vise versa, ensure that the integral is performed over time, and that the corresponding rate is in mergers per unit time (not per unit redshift).

  Here we introduce 3 new sets of initial conditions, designed to bookend the possible parameter space of rates.  In addition to our standard model, in which the distribution of outer orbits runs up to 20,000 AU, we also consider a population synthesis model with a maximum initial $a_2$ of 2,000 AU.  Furthermore, because the kicks are one of the largest uncertainties in many population synthesis studies of BBHs, we also consider models with zero natal kicks for any BHs (though we still treat the effect of the mass loss of the inner binary).  The 4 possible models are presented in Figure \ref{fig:rate}.
  	
    The second critical uncertainty is the possible LK dynamics during the evolution of the stellar triples to BH triples, which we have ignored here.  To bracket this uncertainty, we consider two additional factors: first, for the most pessimistic case, we assume that any
    { stellar} triple whose initial LK timescale is less than the lifetime for the most massive stars ($\sim 3$ Myr) will merge before it evolves to a BH triple.  We then exclude from our sample any triple where $t_{\rm LK} < 3\rm{Myr}$ and Equation \eqref{eqn:suppressed} is false, since a stellar triple may begin its life sufficiently close that pericenter precession suppresses LK oscillations (and possible mergers), then evolve to a regime where LK is possible.  This assumption reduces the overall number of mergers by 90\% to 95\%, but is far more drastic than the results from the triple stellar evolution presented in \cite{Antonini2017}, which found the decrease in surviving BH triples was around 70\% (see e.g., their Figure 2).   For the pessimistic case, we recompute the rate using $t_{\rm delay}$ of those triples that cannot merge as stellar triples.
    
    As is obvious from the figure, the rate increases with the overall SFR of the universe, peaking in around $z \sim 1.6$ for the optimistic cases, and $z \sim 1.2$ for the pessimistic cases.  The delay between the merger rate and the overall peak of low-metallicitiy SFR (at $z\sim 2$) arises from the delay time between formation of stellar triples and the merger of the inner BBH.  The pessimistic case has a peak at lower redshifts, since we have assumed that triples with small LK timescales will merge during the main sequence, which leaves us with a population of mergers with longer delay times. The highest merger rate in any of our models (maximum $a_2$ of 2,000 AU and no BH natal kicks) occurs at $z\sim 1.6$, with a merger rate of nearly $90~\rm{Gpc}^{-3}\rm{yr}^{-1}$, which decreases to $23~\rm{Gpc}^{-3}\rm{yr}^{-1}$ in the local universe ($z\sim 0$). Our most pessimistic case (maximum $a_2$ of 2,000 AU, regular natal kicks, and excluding any systems where $t_{\rm{LK}} < 3\rm{Myr}$) achieves a maximum of $5~\rm{Gpc}^{-3}\rm{yr}^{-1}$ at $z\sim 1.1$, which decreases to $2~\rm{Gpc}^{-3}\rm{yr}^{-1}$ in the local universe.  Combined with the rate of mergers from triples at solar metallicities \citep[$0.3-2.5 ~\rm{Gpc}^{-3}\rm{yr}^{-1}$ in the local universe,][]{Antonini2017}, this suggests an overall merger rate from stellar triples of between 2 and $25~\rm{Gpc}^{-3}\rm{yr}^{-1}$.  Although we do not show the calculation here, our high-metallicity stellar triples ($Z > 0.25Z_{\odot}$) produce a similar merger rate of $0.1-2 ~\rm{Gpc}^{-3}\rm{yr}^{-1}$.  
    
    This range of merger rates is consistent with the current rates from the first Advanced LIGO observing run \citep{Abbott2016e}, but we note that this merger rate applies \emph{only} to the heavy, low-spin BBH mergers detected to date.  Given that these rates are fully consistent with the rate reported by these individual events (2-53 $\rm{Gpc}^{-3}\rm{yr}^{-1}$ for GW150914-type events), we suggest that the merger of stellar triples from low-metallicity environments can naturally explain all the heavy BBHs observed by LIGO/Virgo.
    
\begin{figure}[]
\centering
\includegraphics[scale=0.9]{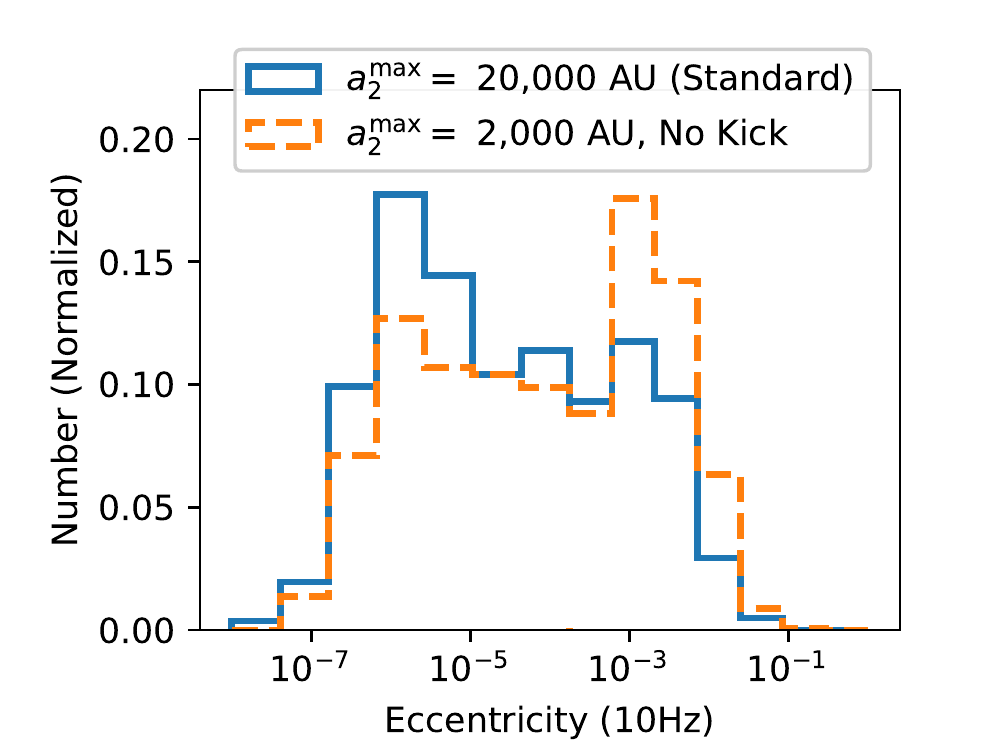}
\caption{The final eccentricity (at a GW frequency of 10Hz) for our standard model (used throughout the main text), and our more liberal model (maximum $a_2$ of 2,000 AU and no BH natal kicks).  While the underlying physics of these two distributions is unchanged (see Sections \ref{sec:2.5} and \ref{subsec:eccspin}), the relative number of systems in each peak (determined by Eqn.~\ref{rqpn}) depends on our initial conditions.}
\label{fig:eccchi2}
\end{figure}

    Finally, we note that these different initial conditions all show similar 
    spin, mass, and eccentricity distributions to those of our standard model 
    that were illustrated in the previous section, with one notable exception.  
    In Figure \ref{fig:eccchi2}, the joint distributions between the eccentricity 
    and spin dependence on the initial LK and GR timescales of the system 
    through Equation \eqref{rqpn}.  While the physics is unchanged, the 
    different initial conditions here populate different regions of this 
    parameter space.  To illustrate this, we show the eccentricity distributions 
    at 10Hz for our standard model (i.e.\ the middle panel of Figure 
    \ref{fig:eccchi2}) and our most liberal model (maximum $a_2$ of 2,000 AU and no BH natal kicks).  While the location of the two peaks are unchanged (as are their respective $\chi_{\rm eff}$ distributions), the relative fraction of sources in each regime of Eqn.~\eqref{rqpn} does.  As such, our somewhat conservative choice of initial conditions used in Section \ref{sec:observable} may be under-predicting the number of sources with large eccentricities and a flat distribution of $\chi_{\rm eff}$.

\section{Conclusion}

In this paper, we have explored the contribution to the GW landscape from BBHs driven to merger by a third BH via the LK mechanism.  By evolving a population of $\sim 10^6$ stellar triples, we found that the reduced mass loss and lower natal kicks for triples at low metallicities ($\lesssim 0.25 Z_{\odot}$) make stellar-mass BH triples nearly 100 times more efficient at producing mergers than triples at stellar metallicities \citep[e.g.,][]{Antonini2017}.  Using self-consistent secular equations for the triple dynamics and BBH spin evolution, we expand upon the results first described in \cite{Antonini2017a}, and show that these low-metallicity-forged triples naturally form heavy BBHs with low effective spins that merge in the local universe, and that the merger rate of these objects ($2-25\rm{Gpc}^{-3}\rm{yr}^{-1}$) can explain all of heavy BBH mergers observed by LIGO/Virgo to date.

Our simplified approach to triple dynamics ignores the potentially significant evolution of stellar triples during their evolution to BH triples.  Although our rate estimate brackets the range of possible mergers and interactions during this phase, a more self-consistent approach \citep[e.g.][]{Toonen2016} will provide further insight into the initial conditions of these systems, since it is likely that our upper-limits have underestimated the number of stellar triples that merged before becoming triple BHs.

At the same time, we have ignored a significant amount of the LK-driven BBH merger parameter space by our assumption that the third body be a BH.  This was done for simplicity (as our technique could not follow LK oscillations while a tertiary was still evolving), but the eccentric LK mechanism does not require the third body to be a BH.  It is entirely possible that, by considering triples in which the inner BH binary is orbited by a distant massive star, that the merger rates quoted here may increase.  

Finally, all of our $\chi_{\rm eff}$ distributions rely on the assumption that the spins of the inner binary are initially aligned with the orbit.  Although this assumption has been used in many previous studies, it is easy to imagine scenarios where this may not be the case, and there exists some observational evidence that ``Binaries are Not Always Neatly Aligned'' \citep[the BANANA survey][]{Albrecht2010,Albrecht2013,Albrecht2014}.  At the same time, many of the inner binaries in the triples explored here experienced significant mass transfer during their main sequence evolution, which could, in conjunction with tidal torques, both spin-up and align the BHs with the orbital angular momentum.  The correct treatment of dynamical tides and eccentric mass transfer during LK oscillations is significantly beyond the scope of this paper, but could play a large role in the initial spin distributions of BH triples.


\begin{acknowledgments}
We would like to thank Michael Kesden, Davide Gerosa, Scott Hughes, Cole Miller, and Richard O'Shaughnessy for useful discussions.  CR acknowledges support from the Pappalardo Fellowship in Physics at MIT.  CR and FA also acknowledge support from NSF Grant PHY-1607611 to the Aspen
Center for Physics.  FA acknowledges support from an E. Rutherford fellowship (ST/P00492X/1) from the Science and Technology Facilities Council.

\end{acknowledgments}

\software{The secular equations derived and used in this paper are available in both C++ and Fortran at \href{https://github.com/carlrodriguez/kozai}{https://github.com/carlrodriguez/kozai}.  They are both free to use, provided that this work and \cite{Antonini2017a} are cited.}

\bibliographystyle{aasjournal}

\begin{thebibliography}{}
\expandafter\ifx\csname natexlab\endcsname\relax\def\natexlab#1{#1}\fi

\bibitem[{Abbott {et~al.}(2016{\natexlab{a}})Abbott, Abbott, Abbott, Abernathy,
  Acernese, Ackley, Adams, Adams, Addesso, Adhikari, Adya, Affeldt, Agathos,
  Agatsuma, Aggarwal, Aguiar, Aiello, Ain, Ajith, Allen, Allocca, Altin,
  Anderson, Anderson, Arai, Araya, Arceneaux, Areeda, Arnaud, Arun, Ascenzi,
  Ashton, Ast, Aston, Astone, Aufmuth, Aulbert, Babak, Bacon, Bader, Baker,
  Baldaccini, Ballardin, Ballmer, Barayoga, Barclay, Barish, Barker, Barone,
  Barr, Barsotti, Barsuglia, Barta, Bartlett, Bartos, Bassiri, Basti, Batch,
  Baune, Bavigadda, Bazzan, Bejger, Bell, Berger, Bergmann, Berry, Bersanetti,
  Bertolini, Betzwieser, Bhagwat, Bhandare, Bilenko, Billingsley, Birch,
  Birney, Birnholtz, Biscans, Bisht, Bitossi, Biwer, Bizouard, Blackburn,
  Blair, Blair, Blair, Bloemen, Bock, Boer, Bogaert, Bogan, Bohe, Bond, Bondu,
  Bonnand, Boom, Bork, Boschi, Bose, Bouffanais, Bozzi, Bradaschia, Brady,
  Braginsky, Branchesi, Brau, Briant, Brillet, Brinkmann, Brisson, Brockill,
  Broida, Brooks, Brown, Brown, Brown, Brunett, Buchanan, Buikema, Bulik,
  Bulten, Buonanno, Buskulic, Buy, Byer, Cabero, Cadonati, Cagnoli, Cahillane,
  Calder{\'{o}}n~Bustillo, Callister, Calloni, Camp, Cannon, Cao, Capano,
  Capocasa, Carbognani, Caride, Casanueva~Diaz, Casentini, Caudill,
  Cavagli{\`{a}}, Cavalier, Cavalieri, Cella, Cepeda, Cerboni~Baiardi,
  Cerretani, Cesarini, Chamberlin, Chan, Chao, Charlton, Chassande-Mottin,
  Cheeseboro, Chen, Chen, Cheng, Chincarini, Chiummo, Cho, Cho, Chow,
  Christensen, Chu, Chua, Chung, Ciani, Clara, Clark, Cleva, Coccia, Cohadon,
  Colla, Collette, Cominsky, Constancio, Conte, Conti, Cook, Corbitt, Cornish,
  Corsi, Cortese, Costa, Coughlin, Coughlin, Coulon, Countryman, Couvares,
  Cowan, Coward, Cowart, Coyne, Coyne, Craig, Creighton, Cripe, Crowder,
  Cumming, Cunningham, Cuoco, Dal~Canton, Danilishin, D’Antonio, Danzmann,
  Darman, Dasgupta, Da~Silva~Costa, Dattilo, Dave, Davier, Davies, Daw, Day,
  De, DeBra, Debreczeni, Degallaix, De~Laurentis, Del{\'{e}}glise, Del~Pozzo,
  Denker, Dent, Dergachev, De~Rosa, DeRosa, DeSalvo, Devine, Dhurandhar,
  D{\'{i}}az, Di~Fiore, Di~Giovanni, Di~Girolamo, Di~Lieto, Di~Pace, Di~Palma,
  Di~Virgilio, Dolique, Donovan, Dooley, Doravari, Douglas, Downes, Drago,
  Drever, Driggers, Ducrot, Dwyer, Edo, Edwards, Effler, Eggenstein, Ehrens,
  Eichholz, Eikenberry, Engels, Essick, Etzel, Evans, Evans, Everett,
  Factourovich, Fafone, Fair, Fairhurst, Fan, Fang, Farinon, Farr, Farr,
  Favata, Fays, Fehrmann, Fejer, Fenyvesi, Ferrante, Ferreira, Ferrini,
  Fidecaro, Fiori, Fiorucci, Fisher, Flaminio, Fletcher, Fong, Fournier,
  Frasca, Frasconi, Frei, Freise, Frey, Frey, Fritschel, Frolov, Fulda, Fyffe,
  Gabbard, Gaebel, Gair, Gammaitoni, Gaonkar, Garufi, Gaur, Gehrels, Gemme,
  Geng, Genin, Gennai, George, Gergely, Germain, Ghosh, Ghosh, Ghosh, Giaime,
  Giardina, Giazotto, Gill, Glaefke, Goetz, Goetz, Gondan, Gonz{\'{a}}lez,
  Gonzalez~Castro, Gopakumar, Gordon, Gorodetsky, Gossan, Gosselin, Gouaty,
  Grado, Graef, Graff, Granata, Grant, Gras, Gray, Greco, Green, Groot, Grote,
  Grunewald, Guidi, Guo, Gupta, Gupta, Gushwa, Gustafson, Gustafson, Hacker,
  Hall, Hall, Hamilton, Hammond, Haney, Hanke, Hanks, Hanna, Hannam, Hanson,
  Hardwick, Harms, Harry, Harry, Hart, Hartman, Haster, Haughian, Healy,
  Heidmann, Heintze, Heitmann, Hello, Hemming, Hendry, Heng, Hennig, Henry,
  Heptonstall, Heurs, Hild, Hoak, Hofman, Holt, Holz, Hopkins, Hough, Houston,
  Howell, Hu, Huang, Huerta, Huet, Hughey, Husa, Huttner, Huynh-Dinh, Indik,
  Ingram, Inta, Isa, Isac, Isi, Isogai, Iyer, Izumi, Jacqmin, Jang, Jani,
  Jaranowski, Jawahar, Jian, Jim{\'{e}}nez-Forteza, Johnson, Johnson-McDaniel,
  Jones, Jones, Jonker, Ju, K, Kalaghatgi, Kalogera, Kandhasamy, Kang, Kanner,
  Kapadia, Karki, Karvinen, Kasprzack, Katsavounidis, Katzman, Kaufer, Kaur,
  Kawabe, K{\'{e}}f{\'{e}}lian, Kehl, Keitel, Kelley, Kells, Kennedy, Key,
  Khalili, Khan, Khan, Khan, Khazanov, Kijbunchoo, Kim, Kim, Kim, Kim, Kim,
  Kim, Kim, Kimbrell, King, King, Kissel, Klein, Kleybolte, Klimenko,
  Koehlenbeck, Koley, Kondrashov, Kontos, Korobko, Korth, Kowalska, Kozak,
  Kringel, Krishnan, Kr{\'{o}}lak, Krueger, Kuehn, Kumar, Kumar, Kuo, Kutynia,
  Lackey, Landry, Lange, Lantz, Lasky, Laxen, Lazzarini, Lazzaro, Leaci,
  Leavey, Lebigot, Lee, Lee, Lee, Lee, Lenon, Leonardi, Leong, Leroy, Letendre,
  Levin, Lewis, Li, Libson, Littenberg, Lockerbie, Lombardi, London, Lord,
  Lorenzini, Loriette, Lormand, Losurdo, Lough, Lousto, L{\"{u}}ck, Lundgren,
  Lynch, Ma, Machenschalk, MacInnis, Macleod, Maga{\~{n}}a-Sandoval,
  Maga{\~{n}}a~Zertuche, Magee, Majorana, Maksimovic, Malvezzi, Man, Mandel,
  Mandic, Mangano, Mansell, Manske, Mantovani, Marchesoni, Marion, M{\'{a}}rka,
  M{\'{a}}rka, Markosyan, Maros, Martelli, Martellini, Martin, Martynov, Marx,
  Mason, Masserot, Massinger, Masso-Reid, Mastrogiovanni, Matichard, Matone,
  Mavalvala, Mazumder, McCarthy, McClelland, McCormick, McGuire, McIntyre,
  McIver, McManus, McRae, McWilliams, Meacher, Meadors, Meidam, Melatos,
  Mendell, Mercer, Merilh, Merzougui, Meshkov, Messenger, Messick, Metzdorff,
  Meyers, Mezzani, Miao, Michel, Middleton, Mikhailov, Milano, Miller, Miller,
  Miller, Miller, Millhouse, Minenkov, Ming, Mirshekari, Mishra, Mitra,
  Mitrofanov, Mitselmakher, Mittleman, Moggi, Mohan, Mohapatra, Montani, Moore,
  Moore, Moraru, Moreno, Morriss, Mossavi, Mours, Mow-Lowry, Mueller, Muir,
  Mukherjee, Mukherjee, Mukherjee, Mukund, Mullavey, Munch, Murphy, Murray,
  Mytidis, Nardecchia, Naticchioni, Nayak, Nedkova, Nelemans, Nelson, Neri,
  Neunzert, Newton, Nguyen, Nielsen, Nissanke, Nitz, Nocera, Nolting,
  Normandin, Nuttall, Oberling, Ochsner, O’Dell, Oelker, Ogin, Oh, Oh, Ohme,
  Oliver, Oppermann, Oram, O’Reilly, O’Shaughnessy, Ottaway, Overmier,
  Owen, Pai, Pai, Palamos, Palashov, Palomba, Pal-Singh, Pan, Pan, Pankow,
  Pannarale, Pant, Paoletti, Paoli, Papa, Paris, Parker, Pascucci, Pasqualetti,
  Passaquieti, Passuello, Patricelli, Patrick, Pearlstone, Pedraza, Pedurand,
  Pekowsky, Pele, Penn, Perreca, Perri, Pfeiffer, Phelps, Piccinni, Pichot,
  Piergiovanni, Pierro, Pillant, Pinard, Pinto, Pitkin, Poe, Poggiani,
  Popolizio, Porter, Post, Powell, Prasad, Predoi, Prestegard, Price,
  Prijatelj, Principe, Privitera, Prix, Prodi, Prokhorov, Puncken, Punturo,
  Puppo, P{\"{u}}rrer, Qi, Qin, Qiu, Quetschke, Quintero, Quitzow-James, Raab,
  Rabeling, Radkins, Raffai, Raja, Rajan, Rakhmanov, Rapagnani, Raymond,
  Razzano, Re, Read, Reed, Regimbau, Rei, Reid, Reitze, Rew, Reyes, Ricci,
  Riles, Rizzo, Robertson, Robie, Robinet, Rocchi, Rolland, Rollins, Roma,
  Romano, Romano, Romanov, Romie, Rosi{\'{n}}ska, Rowan, R{\"{u}}diger, Ruggi,
  Ryan, Sachdev, Sadecki, Sadeghian, Sakellariadou, Salconi, Saleem, Salemi,
  Samajdar, Sammut, Sanchez, Sandberg, Sandeen, Sanders, Sassolas,
  Sathyaprakash, Saulson, Sauter, Savage, Sawadsky, Schale, Schilling, Schmidt,
  Schmidt, Schnabel, Schofield, Sch{\"{o}}nbeck, Schreiber, Schuette, Schutz,
  Scott, Scott, Sellers, Sengupta, Sentenac, Sequino, Sergeev, Setyawati,
  Shaddock, Shaffer, Shahriar, Shaltev, Shapiro, Shawhan, Sheperd, Shoemaker,
  Shoemaker, Siellez, Siemens, Sieniawska, Sigg, Silva, Singer, Singer, Singh,
  Singh, Singhal, Sintes, Slagmolen, Smith, Smith, Smith, Son, Sorazu,
  Sorrentino, Souradeep, Srivastava, Staley, Steinke, Steinlechner,
  Steinlechner, Steinmeyer, Stephens, Stevenson, Stone, Strain, Straniero,
  Stratta, Strauss, Strigin, Sturani, Stuver, Summerscales, Sun, Sunil, Sutton,
  Swinkels, Szczepa{\'{n}}czyk, Tacca, Talukder, Tanner, T{\'{a}}pai, Tarabrin,
  Taracchini, Taylor, Theeg, Thirugnanasambandam, Thomas, Thomas, Thomas,
  Thorne, Thrane, Tiwari, Tiwari, Tokmakov, Toland, Tomlinson, Tonelli,
  Tornasi, Torres, Torrie, T{\"{o}}yr{\"{a}}, Travasso, Traylor, Trifir{\`{o}},
  Tringali, Trozzo, Tse, Turconi, Tuyenbayev, Ugolini, Unnikrishnan, Urban,
  Usman, Vahlbruch, Vajente, Valdes, Vallisneri, van Bakel, van Beuzekom,
  van~den Brand, Van Den~Broeck, Vander-Hyde, van~der Schaaf, van Heijningen,
  van Veggel, Vardaro, Vass, Vas{\'{u}}th, Vaulin, Vecchio, Vedovato, Veitch,
  Veitch, Venkateswara, Verkindt, Vetrano, Vicer{\'{e}}, Vinciguerra, Vine,
  Vinet, Vitale, Vo, Vocca, Vorvick, Voss, Vousden, Vyatchanin, Wade, Wade,
  Wade, Walker, Wallace, Walsh, Wang, Wang, Wang, Wang, Wang, Ward, Warner,
  Was, Weaver, Wei, Weinert, Weinstein, Weiss, Wen, We{\ss}els, Westphal,
  Wette, Whelan, Whitcomb, Whiting, Williams, Williamson, Willis, Willke,
  Wimmer, Winkler, Wipf, Wittel, Woan, Woehler, Worden, Wright, Wu, Wu, Yablon,
  Yam, Yamamoto, Yancey, Yu, Yvert, Zadro{\.{z}}ny, Zangrando, Zanolin, Zendri,
  Zevin, Zhang, Zhang, Zhang, Zhao, Zhou, Zhou, Zhu, Zucker, Zuraw, \&
  Zweizig}]{Abbott2016e}
Abbott, B.~P., Abbott, R., Abbott, T.~D., {et~al.} 2016{\natexlab{a}}, PRX, 6, 041015

\bibitem[{Abbott {et~al.}(2016{\natexlab{b}})Abbott, Abbott, Abbott, Abernathy,
  Acernese, Ackley, Adams, Adams, Addesso, Adhikari, Adya, \&
  Affeldt}]{Abbott2016}
---. 2016{\natexlab{b}}, PRL, 116, 241103

\bibitem[{Abbott {et~al.}(2016{\natexlab{c}})Abbott, Abbott, Abbott, Abernathy,
  Acernese, Ackley, Adams, Adams, Addesso, Adhikari, \& Adya}]{Abbott2016a}
---. 2016{\natexlab{c}}, PRL, 116, 061102

\bibitem[{Abbott {et~al.}(2017{\natexlab{a}})Abbott, Abbott, Abbott, Acernese,
  Ackley, Adams, Adams, Addesso, Adhikari, \& Adya}]{Abbott2017}
---. 2017{\natexlab{a}}, PRL, 118, 221101

\bibitem[{Abbott {et~al.}(2017{\natexlab{b}})Abbott, Abbott, Abbott, Acernese,
  Ackley, Adams, Adams, Addesso, Adhikari, \& Adya}]{Abbott2017e}
---. 2017{\natexlab{b}}, ApJ, 851, L35

\bibitem[{Abbott {et~al.}(2017{\natexlab{c}})Abbott, Abbott, Abbott, Acernese,
  Ackley, Adams, Adams, Addesso, Adhikari, \& Adya}]{Abbott2017c}
---. 2017{\natexlab{c}}, PRL, 119, 1

\bibitem[{Ajith {et~al.}(2011)Ajith, Hannam, Husa, Chen, Br{\"{u}}gmann,
  Dorband, M{\"{u}}ller, Ohme, Pollney, Reisswig, Santamar{\'{i}}a, \&
  Seiler}]{Ajith2011}
Ajith, P., Hannam, M., Husa, S., {et~al.} 2011, PRL, 106, 6

\bibitem[{Albrecht {et~al.}(2013)Albrecht, Setiawan, Torres, Fabrycky, \&
  Winn}]{Albrecht2013}
Albrecht, S., Setiawan, J., Torres, G., Fabrycky, D.~C., \& Winn, J.~N. 2013,
  ApJ, 767, 1

\bibitem[{Albrecht {et~al.}(2010)Albrecht, Winn, Carter, Snellen, \&
  de~Mooij}]{Albrecht2010}
Albrecht, S., Winn, J., Carter, J., Snellen, I., \& de~Mooij, E. 2010, ApJ, 726, 2

\bibitem[{Albrecht {et~al.}(2014)Albrecht, Winn, Torres, Fabrycky, Setiawan,
  Gillon, Jehin, Triaud, Queloz, Snellen, \& Eggleton}]{Albrecht2014}
Albrecht, S., Winn, J.~N., Torres, G., {et~al.} 2014, ApJ, 785, 83

\bibitem[{Antonini {et~al.}(2014)Antonini, Murray, \& Mikkola}]{Antonini2014}
Antonini, F., Murray, N., \& Mikkola, S. 2014, ApJ, 781,
  45

\bibitem[{Antonini \& Perets(2012)}]{Antonini2012a}
Antonini, F., \& Perets, H.~B. 2012, ApJ, 757, 27

\bibitem[{Antonini \& Rasio(2016)}]{Antonini2016}
Antonini, F., \& Rasio, F.~A. 2016, ApJ, 231, 187

\bibitem[{Antonini {et~al.}(2017{\natexlab{a}})Antonini, Rodriguez, Petrovich,
  \& Fischer}]{Antonini2017a}
Antonini, F., Rodriguez, C.~L., Petrovich, C., \& Fischer, C.~L.
  2017{\natexlab{a}}, astro-ph:1711.07142

\bibitem[{Antonini {et~al.}(2017{\natexlab{b}})Antonini, Toonen, \&
  Hamers}]{Antonini2017}
Antonini, F., Toonen, S., \& Hamers, A.~S. 2017{\natexlab{b}}, ApJ, 841, 77

\bibitem[{Askar {et~al.}(2016)Askar, Szkudlarek, Gondek-Rosi{\'{n}}ska, Giersz,
  \& Bulik}]{Askar2016}
Askar, A., Szkudlarek, M., Gondek-Rosi{\'{n}}ska, D., Giersz, M., \& Bulik, T.
  2016, MNRAS, 464, L36

\bibitem[{Bae {et~al.}(2014)Bae, Kim, \& Lee}]{Bae2014}
Bae, Y.-B., Kim, C., \& Lee, H.~M. 2014, MNRAS, 440, 2714

\bibitem[{Banerjee(2017)}]{Banerjee2017}
Banerjee, S. 2017, MNRAS, 467, 524

\bibitem[{Banerjee {et~al.}(2010)Banerjee, Baumgardt, \& Kroupa}]{Banerjee2010}
Banerjee, S., Baumgardt, H., \& Kroupa, P. 2010, MNRAS, 402, 371

\bibitem[{Barker \& O'Connell(1975)}]{Barker1975}
Barker, B.~M., \& O'Connell, R.~F. 1975, PRD, 12, 329

\bibitem[{Bartos {et~al.}(2016)Bartos, Kocsis, Haiman, \&
  M{\'{a}}rka}]{Bartos2016}
Bartos, I., Kocsis, B., Haiman, Z., \& M{\'{a}}rka, S. 2016, 1

\bibitem[{Belczynski {et~al.}(2010)Belczynski, Dominik, Bulik, O’Shaughnessy,
  Fryer, \& Holz}]{Belczynski2010}
Belczynski, K., Dominik, M., Bulik, T., {et~al.} 2010, ApJ, 715, L138

\bibitem[{Belczynski {et~al.}(2016)Belczynski, Holz, Bulik, \&
  O’Shaughnessy}]{Belczynski2016}
Belczynski, K., Holz, D.~E., Bulik, T., \& O’Shaughnessy, R. 2016, Nature,
  534, 512

\bibitem[{Belczynski {et~al.}(2002)Belczynski, Kalogera, \&
  Bulik}]{Belczynski2002}
Belczynski, K., Kalogera, V., \& Bulik, T. 2002, ApJ,
  572, 407

\bibitem[{Blaes {et~al.}(2002)Blaes, Lee, \& Socrates}]{Blaes2002}
Blaes, O., Lee, M.~H., \& Socrates, A. 2002, ApJ, 578,
  775

\bibitem[{Breivik {et~al.}(2016)Breivik, Rodriguez, Larson, Kalogera, Rasio,
  Abbott B.~P., Abbott B.~P., Abbott B.~P., Antonini~F., Antonini~F., S.,
  Belczynski~K., T., Belczynski~K., R., Belczynski~K., T., Bell E.~F., D.,
  Bird~S., Chatterjee~S., A., Correnti~M., A., Dominik~M., A., F., Fryer C.~L.,
  V., Kalogera, T., Mazeh, Z., E., Harris W.~E., C., A., Hobbs~G., M., Hurley
  J.~R., R., Kobulnicky H.~A., P., E., de~Mink~S., Marchant~P., J., Mazeh~T.,
  M., Morscher~M., A., Nishizawa~A., A., Pattabiraman~B., C., R., L., Rodriguez
  C.~L., A., Rodriguez C.~L., A., Samsing~J., E., A., N., N., Seto, A.,
  de~Koter, Vink J.~S., M., Vitale~S., P., \& L.}]{Breivik2016}
Breivik, K., Rodriguez, C.~L., Larson, S.~L., {et~al.} 2016, ApJ, 830, L18

\bibitem[{Buonanno {et~al.}(2011)Buonanno, Kidder, Mrou{\'{e}}, Pfeiffer, \&
  Taracchini}]{Buonanno2011}
Buonanno, A., Kidder, L.~E., Mrou{\'{e}}, A.~H., Pfeiffer, H.~P., \&
  Taracchini, A. 2011, PRD, 83, 1

\bibitem[{Correia {et~al.}(2011)Correia, Laskar, Farago, \&
  Bou{\'{e}}}]{Correia2011}
Correia, A.~C., Laskar, J., Farago, F., \& Bou{\'{e}}, G. 2011, Celestial
  Mechanics and Dynamical Astronomy, 111, 105

\bibitem[{Damour(2001)}]{Damour2001}
Damour, T. 2001, PRD, 64, 22

\bibitem[{Damour \& Sch{\"{a}}eer(1988)}]{Damour1988}
Damour, T., \& Sch{\"{a}}eer, G. 1988, Il Nuovo Cimento B, 101, 127

\bibitem[{Davies(2017)}]{Davies2017}
Davies, M. 2017, Talk at the Kavli Summer Program in Astrophysics, Copenhagen, 2017

\bibitem[{De~Mink \& Mandel(2016)}]{DeMink2016}
De~Mink, S.~E., \& Mandel, I. 2016, MNRAS, 460, 3545

\bibitem[{Dominik {et~al.}(2012)Dominik, Belczynski, Fryer, Holz, Berti, Bulik,
  Mandel, \& O'Shaughnessy}]{Dominik2012}
Dominik, M., Belczynski, K., Fryer, C., {et~al.} 2012, ApJ, 759, 52

\bibitem[{Dominik {et~al.}(2013)Dominik, Belczynski, Fryer, Holz, Berti, Bulik,
  Mandel, \& O'Shaughnessy}]{Dominik2013}
---. 2013, ApJ, 779, 72

\bibitem[{Dominik {et~al.}(2015)Dominik, Berti, O’Shaughnessy, Mandel,
  Belczynski, Fryer, Holz, Bulik, \& Pannarale}]{Dominik2014}
Dominik, M., Berti, E., O’Shaughnessy, R., {et~al.} 2015, ApJ, 806, 263

\bibitem[{Downing {et~al.}(2010)Downing, Benacquista, Giersz, \&
  Spurzem}]{Downing2010}
Downing, J. M.~B., Benacquista, M.~J., Giersz, M., \& Spurzem, R. 2010, MNRAS, 407, 1946

\bibitem[{Downing {et~al.}(2011)Downing, Benacquista, Giersz, \&
  Spurzem}]{Downing2011}
---. 2011, MNRAS, 416, 133

\bibitem[{Dvorkin {et~al.}(2015)Dvorkin, Silk, Vangioni, Petitjean, \&
  Olive}]{Dvorkin2015}
Dvorkin, I., Silk, J., Vangioni, E., Petitjean, P., \& Olive, K.~A. 2015,
  MNRAS, 452, L36

\bibitem[{Eggleton \& Kiseleva-‐Eggleton(2001)}]{Eggleton2001}
Eggleton, P., \& Kiseleva-‐Eggleton, L. 2001, ApJ,
  562, 1012

\bibitem[{Fabrycky \& Tremaine(2007)}]{Fabrycky2007}
Fabrycky, D., \& Tremaine, S. 2007, ApJ, 669, 1298

\bibitem[{Farr {et~al.}(2018)Farr, Holz, \& Farr}]{Farr2018}
Farr, B., Holz, D.~E., \& Farr, W.~M. 2018, ApJ, 854, L9

\bibitem[{Farr {et~al.}(2017)Farr, Stevenson, Miller, Mandel, Farr, \&
  Vecchio}]{Farr2017}
Farr, W.~M., Stevenson, S., Miller, M.~C., {et~al.} 2017, Nature, 548, 426

\bibitem[{Ford {et~al.}(2000)Ford, Kozinsky, \& Rasio}]{Ford2000}
Ford, E.~B., Kozinsky, B., \& Rasio, F.~A. 2000, ApJ,
  535, 385

\bibitem[{Fryer {et~al.}(2012)Fryer, Belczynski, Wiktorowicz, Dominik,
  Kalogera, \& Holz}]{Fryer2012}
Fryer, C.~L., Belczynski, K., Wiktorowicz, G., {et~al.} 2012, ApJ, 749, 91

\bibitem[{Fryer \& Kalogera(2001)}]{Fryer2001}
Fryer, C.~L., \& Kalogera, V. 2001, ApJ, 554, 548

\bibitem[{Gerosa {et~al.}(2013)Gerosa, Kesden, Berti, O’Shaughnessy, \&
  Sperhake}]{Gerosa2013}
Gerosa, D., Kesden, M., Berti, E., O’Shaughnessy, R., \& Sperhake, U. 2013,
  PRD, 87, 104028

\bibitem[{Gerosa {et~al.}(2014)Gerosa, O'Shaughnessy, Kesden, Berti, \&
  Sperhake}]{Gerosab}
Gerosa, D., O'Shaughnessy, R., Kesden, M., Berti, E., \& Sperhake, U. 2014,
  PRD, 89, 124025

\bibitem[{Giesler {et~al.}(2018)Giesler, Clausen, \& Ott}]{Giesler2017}
Giesler, M., Clausen, D., \& Ott, C.~D. 2018, MNRAS, 477, 1853

\bibitem[{Harrington(1968)}]{Harrington1968}
Harrington, R.~S. 1968, AJ, 73, 190

\bibitem[{Hoang {et~al.}(2018)Hoang, Naoz, Kocsis, Rasio, \&
  Dosopoulou}]{Hoang2018}
Hoang, B.-M., Naoz, S., Kocsis, B., Rasio, F.~A., \& Dosopoulou, F. 2018, ApJ, 856, 140

\bibitem[{Hurley {et~al.}(2000)Hurley, Pols, \& Tout}]{Hurley2000}
Hurley, J.~R., Pols, O.~R., \& Tout, C.~A. 2000, MNRAS, 315, 543

\bibitem[{Hurley {et~al.}(2002)Hurley, Tout, \& Pols}]{Hurley2002}
Hurley, J.~R., Tout, C.~A., \& Pols, O.~R. 2002, MNRAS, 329, 897

\bibitem[{Kalogera(2000)}]{Kalogera2000}
Kalogera, V. 2000, ApJ, 541, 319

\bibitem[{Kozai(1962)}]{Kozai1962}
Kozai, Y. 1962, ApJ, 67, 591

\bibitem[{Kroupa(2001)}]{Kroupa2001}
Kroupa, P. 2001, MNRAS, 322, 231

\bibitem[{Leigh {et~al.}(2018)Leigh, Geller, McKernan, Ford, Mac~Low,
  Bellovary, Haiman, Lyra, Samsing, O'Dowd, Kocsis, \& Endlich}]{Leigh2017}
Leigh, N. W.~C., Geller, A.~M., McKernan, B., {et~al.} 2018, MNRAS, 474, 5672

\bibitem[{Lidov(1962)}]{Lidov1962}
Lidov, M. 1962, Planetary and Space Science, 9, 719

\bibitem[{Liu \& Lai(2017)}]{Liu2017}
Liu, B., \& Lai, D. 2017, ApJ, 846, L11

\bibitem[{Liu \& Lai(2018)}]{Liu2018}
---. 2018, astro-ph: 1805.03202

\bibitem[{Liu {et~al.}(2015)Liu, Mu{\~{n}}oz, \& Lai}]{Liu2015}
Liu, B., Mu{\~{n}}oz, D.~J., \& Lai, D. 2015, MNRAS, 447, 747

\bibitem[{Madau \& Dickinson(2014)}]{Madau2014}
Madau, P., \& Dickinson, M. 2014, 415

\bibitem[{Mandel \& De~Mink(2016)}]{Mandel2016a}
Mandel, I., \& De~Mink, S.~E. 2016, MNRAS, 458, 2634

\bibitem[{Marchant {et~al.}(2016)Marchant, Langer, Podsiadlowski, Tauris, \&
  Moriya}]{Marchant2016}
Marchant, P., Langer, N., Podsiadlowski, P., Tauris, T.~M., \& Moriya, T.~J.
  2016, A{\&}A, 588, A50

\bibitem[{Mardling \& Aarseth(2001)}]{Mardling2001}
Mardling, R.~A., \& Aarseth, S.~J. 2001, MNRAS, 321, 398

\bibitem[{Merritt(2013)}]{Merritt2013}
Merritt, D. 2013, {Dynamics and Evolution of Galactic Nuclei} (Princeton:
  Princeton University Press), 544

\bibitem[{Miller \& Lauburg(2009)}]{Miller2009}
Miller, M.~C., \& Lauburg, V.~M. 2009, ApJ, 692, 917

\bibitem[{Moody \& Sigurdsson(2009)}]{Moody2009}
Moody, K., \& Sigurdsson, S. 2009, ApJ, 690, 1370

\bibitem[{Naoz(2016)}]{Naoz2016}
Naoz, S. 2016, ARAA, 54, 441

\bibitem[{Naoz {et~al.}(2011)Naoz, Farr, Lithwick, Rasio, \&
  Teyssandier}]{Naoz2011}
Naoz, S., Farr, W.~M., Lithwick, Y., Rasio, F.~A., \& Teyssandier, J. 2011,
  Nature, 473, 187

\bibitem[{Naoz {et~al.}(2013)Naoz, Farr, Lithwick, Rasio, \&
  Teyssandier}]{Naoz2013}
---. 2013, MNRAS, 431, 2155

\bibitem[{O'Leary {et~al.}(2009)O'Leary, Kocsis, \& Loeb}]{Oleary2009}
O'Leary, R.~M., Kocsis, B., \& Loeb, A. 2009, MNRAS, 395, 2127

\bibitem[{O’Leary {et~al.}(2007)O’Leary, O’Shaughnessy, \&
  Rasio}]{OLeary2007}
O’Leary, R., O’Shaughnessy, R., \& Rasio, F. 2007, PRD, 76,
  061504

\bibitem[{O’Leary {et~al.}(2006)O’Leary, Rasio, Fregeau, Ivanova, \&
  O’Shaughnessy}]{OLeary2006}
O’Leary, R.~M., Rasio, F.~A., Fregeau, J.~M., Ivanova, N., \&
  O’Shaughnessy, R. 2006, ApJ, 637, 937

\bibitem[{Peters(1964)}]{Peters1964}
Peters, P. 1964, Physical Review, 136, B1224

\bibitem[{Petrovich(2015)}]{Petrovich2015}
Petrovich, C. 2015, ApJ, 799, 27

\bibitem[{Petrovich \& Antonini(2017)}]{Petrovich2017}
Petrovich, C., \& Antonini, F. 2017, ApJ, 846, 146

\bibitem[{{Ade} {et~al.}(2015){Planck Collaboration}, Ade,
  Aghanim, Arnaud, Ashdown, Aumont, Baccigalupi, Banday, Barreiro, Bartlett,
  Bartolo, Battaner, Battye, Benabed, Benoit, Benoit-Levy, Bernard, Bersanelli,
  Bielewicz, Bonaldi, Bonavera, Bond, Borrill, Bouchet, Boulanger, Bucher,
  Burigana, Butler, Calabrese, Cardoso, Catalano, Challinor, Chamballu, Chary,
  Chiang, Chluba, Christensen, Church, Clements, Colombi, Colombo, Combet,
  Coulais, Crill, Curto, Cuttaia, Danese, Davies, Davis, de~Bernardis, de~Rosa,
  de~Zotti, Delabrouille, Desert, Di~Valentino, Dickinson, Diego, Dolag, Dole,
  Donzelli, Dore, Douspis, Ducout, Dunkley, Dupac, Efstathiou, Elsner, Ensslin,
  Eriksen, Farhang, Fergusson, Finelli, Forni, Frailis, Fraisse, Franceschi,
  Frejsel, Galeotta, Galli, Ganga, Gauthier, Gerbino, Ghosh, Giard,
  Giraud-Heraud, Giusarma, Gjerlow, Gonzalez-Nuevo, Gorski, Gratton, Gregorio,
  Gruppuso, Gudmundsson, Hamann, Hansen, Hanson, Harrison, Helou,
  Henrot-Versille, Hernandez-Monteagudo, Herranz, Hildebrandt, Hivon, Hobson,
  Holmes, Hornstrup, Hovest, Huang, Huffenberger, Hurier, Jaffe, Jaffe, Jones,
  Juvela, Keihanen, Keskitalo, Kisner, Kneissl, Knoche, Knox, Kunz,
  Kurki-Suonio, Lagache, Lahteenmaki, Lamarre, Lasenby, Lattanzi, Lawrence,
  Leahy, Leonardi, Lesgourgues, Levrier, Lewis, Liguori, Lilje, Linden-Vornle,
  Lopez-Caniego, Lubin, Macias-Perez, Maggio, Maino, Mandolesi, Mangilli,
  Marchini, Martin, Martinelli, Martinez-Gonzalez, Masi, Matarrese, Mazzotta,
  McGehee, Meinhold, Melchiorri, Melin, Mendes, Mennella, Migliaccio, Millea,
  Mitra, Miville-Deschenes, Moneti, Montier, Morgante, Mortlock, Moss, Munshi,
  Murphy, Naselsky, Nati, Natoli, Netterfield, Norgaard-Nielsen, Noviello,
  Novikov, Novikov, Oxborrow, Paci, Pagano, Pajot, Paladini, Paoletti,
  Partridge, Pasian, Patanchon, Pearson, Perdereau, Perotto, Perrotta,
  Pettorino, Piacentini, Piat, Pierpaoli, Pietrobon, Plaszczynski,
  Pointecouteau, Polenta, Popa, Pratt, Prezeau, Prunet, Puget, Rachen, Reach,
  Rebolo, Reinecke, Remazeilles, Renault, Renzi, Ristorcelli, Rocha, Rosset,
  Rossetti, Roudier, D'Orfeuil, Rowan-Robinson, Rubino-Martin, Rusholme, Said,
  Salvatelli, Salvati, Sandri, Santos, Savelainen, Savini, Scott, Seiffert,
  Serra, Shellard, Spencer, Spinelli, Stolyarov, Stompor, Sudiwala, Sunyaev,
  Sutton, Suur-Uski, Sygnet, Tauber, Terenzi, Toffolatti, Tomasi, Tristram,
  Trombetti, Tucci, Tuovinen, Turler, Umana, Valenziano, Valiviita, Van~Tent,
  Vielva, Villa, Wade, Wandelt, Wehus, White, White, Wilkinson, Yvon, Zacchei,
  \& Zonca}]{PlanckCollaboration2015}
Ade, P. A.~R., Aghanim, N., Arnaud, M. {et~al.} 2016, A\&A 594, A13

\bibitem[{Podsiadlowski {et~al.}(2003)Podsiadlowski, Rappaport, \&
  Han}]{Podsiadlowski2003}
Podsiadlowski, P., Rappaport, S., \& Han, Z. 2003, MNRAS,
  341, 385

\bibitem[{Portegies~Zwart \& Mcmillan(2000)}]{PortegiesZwart2000}
Portegies~Zwart, S.~F., \& Mcmillan, S. L.~W. 2000, ApJ,
  528, 17

\bibitem[{P{\"{u}}rrer {et~al.}(2016)P{\"{u}}rrer, Hannam, \&
  Ohme}]{Purrer2016}
P{\"{u}}rrer, M., Hannam, M., \& Ohme, F. 2016, PRD, 93, 1

\bibitem[{Repetto \& Nelemans(2015)}]{Repetto2015}
Repetto, S., \& Nelemans, G. 2015, MNRAS, 453, 3341

\bibitem[{Rodriguez {et~al.}(2018)Rodriguez, Amaro-Seoane, Chatterjee, \&
  Rasio}]{Rodriguez2018}
Rodriguez, C.~L., Amaro-Seoane, P., Chatterjee, S., \& Rasio, F.~A. 2018,
  PRL, 120, 151101

\bibitem[{Rodriguez {et~al.}(2016{\natexlab{a}})Rodriguez, Chatterjee, \&
  Rasio}]{Rodriguez2016a}
Rodriguez, C.~L., Chatterjee, S., \& Rasio, F.~A. 2016{\natexlab{a}}, PRD, 93, 084029

\bibitem[{Rodriguez {et~al.}(2016{\natexlab{b}})Rodriguez, Haster, Chatterjee,
  Kalogera, \& Rasio}]{Rodriguez2016b}
Rodriguez, C.~L., Haster, C.-J., Chatterjee, S., Kalogera, V., \& Rasio, F.~A.
  2016{\natexlab{b}}, ApJ, 824, L8

\bibitem[{Rodriguez {et~al.}(2015)Rodriguez, Morscher, Pattabiraman,
  Chatterjee, Haster, \& Rasio}]{Rodriguez2015a}
Rodriguez, C.~L., Morscher, M., Pattabiraman, B., {et~al.} 2015, PRL, 115, 051101

\bibitem[{Rodriguez {et~al.}(2016{\natexlab{c}})Rodriguez, Zevin, Pankow,
  Kalogera, \& Rasio}]{Rodriguez2016c}
Rodriguez, C.~L., Zevin, M., Pankow, C., Kalogera, V., \& Rasio, F.~A.
  2016{\natexlab{c}}, ApJ, 832, L2

\bibitem[{Sadowski {et~al.}(2008)Sadowski, Belczynski, Bulik, Ivanova, Rasio,
  \& O’Shaughnessy}]{Sadowski2007a}
Sadowski, A., Belczynski, K., Bulik, T., {et~al.} 2008, ApJ, 676, 1162

\bibitem[{Sana {et~al.}(2012)Sana, de~Mink, de~Koter, Langer, Evans, Gieles,
  Gosset, Izzard, Le~Bouquin, \& Schneider}]{Sana2012}
Sana, H., de~Mink, S.~E., de~Koter, A., {et~al.} 2012, Science, 337, 444

\bibitem[{Sana {et~al.}(2014)Sana, Le~Bouquin, Lacour, Berger, Duvert, Gauchet,
  Norris, Olofsson, Pickel, Zins, Absil, De~Koter, Kratter, Schnurr, \&
  Zinnecker}]{Sana2014}
Sana, H., Le~Bouquin, J.~B., Lacour, S., {et~al.} 2014, ApJSS, 215, 15

\bibitem[{Schnittman(2004)}]{Schnittman2004}
Schnittman, J.~D. 2004, PRD, 70, 12

\bibitem[{Silsbee \& Tremaine(2017)}]{Silsbee2017}
Silsbee, K., \& Tremaine, S. 2017, ApJ, 836, 39

\bibitem[{Stone {et~al.}(2017)Stone, Metzger, \& Haiman}]{Stone2016}
Stone, N.~C., Metzger, B.~D., \& Haiman, Z. 2017, MNRAS, 464, 1

\bibitem[{Storch {et~al.}(2014)Storch, Anderson, \& Lai}]{Storch2014}
Storch, N.~I., Anderson, K.~R., \& Lai, D. 2014, Science, 345, 1317

\bibitem[{Storch \& Lai(2015)}]{Storch2015}
Storch, N.~I., \& Lai, D. 2015, MNRAS, 448, 1821

\bibitem[{Tanikawa(2013)}]{Tanikawa2013}
Tanikawa, A. 2013, MNRAS, 435, 1358

\bibitem[{Toonen {et~al.}(2016)Toonen, Hamers, \& Zwart}]{Toonen2016}
Toonen, S., Hamers, A., \& Zwart, S.~P. 2016, Computational Astrophysics and
  Cosmology, 3, 6

\bibitem[{Tremaine {et~al.}(2009)Tremaine, Touma, \& Namouni}]{Tremaine2009}
Tremaine, S., Touma, J., \& Namouni, F. 2009, ApJ, 137,
  3706

\bibitem[{Tremaine \& Yavetz(2014)}]{Tremaine2014}
Tremaine, S., \& Yavetz, T.~D. 2014, American Journal of Physics, 82, 769

\bibitem[{Trifir{\`{o}} {et~al.}(2016)Trifir{\`{o}}, O'Shaughnessy, Gerosa,
  Berti, Kesden, Littenberg, \& Sperhake}]{Trifiro}
Trifir{\`{o}}, D., O'Shaughnessy, R., Gerosa, D., {et~al.} 2016, PRD, 93, 4

\bibitem[{VanLandingham {et~al.}(2016)VanLandingham, Miller, Hamilton, \&
  Richardson}]{VanLandingham2016}
VanLandingham, J.~H., Miller, M.~C., Hamilton, D.~P., \& Richardson, D.~C.
  2016, ApJ, 828, 77

\bibitem[{Vink {et~al.}(2001)Vink, de~Koter, \& Lamers}]{Vink2001}
Vink, J.~S., de~Koter, A., \& Lamers, H. J. G. L.~M. 2001, A\&A, 369, 574

\bibitem[{Vitale {et~al.}(2014)Vitale, Lynch, Veitch, Raymond, \&
  Sturani}]{Vitale2014}
Vitale, S., Lynch, R., Veitch, J., Raymond, V., \& Sturani, R. 2014, PRL, 112, 1

\bibitem[{Vitale {et~al.}(2017)Vitale, Lynch, Sturani, \& Graff}]{Vitale2017}
Vitale, S., Lynch, R., Sturani, R. \& Gradd, P. 2017, CQG, 34, 3

\bibitem[{Voss \& Tauris(2003)}]{Voss2003}
Voss, R., \& Tauris, T.~M. 2003, MNRAS, 342, 1169

\bibitem[{Wen(2003)}]{Wen2003}
Wen, L. 2003, ApJ, 598, 419

\bibitem[{Ziosi {et~al.}(2014)Ziosi, Mapelli, Branchesi, \& Tormen}]{Ziosi2014}
Ziosi, B.~M., Mapelli, M., Branchesi, M., \& Tormen, G. 2014, MNRAS, 441, 3703

\end{thebibliography}

\end{document}